\documentclass[amssymb,amsmath,preprint,epsfig,graphics]{revtex4}
\usepackage{epsfig,amssymb,amsmath}
\newcommand{\bs}{\,-\,}
\newcommand{\kl}{(\,}
\newcommand{\kr}{\,)}
\newcommand{\rf}[1]{(\,\ref{#1}\,)}

\newcommand{\pmu}{\partial_{\mu}}

\newcommand{\nmu}{\nabla_{\mu}}
\newcommand{\nmo}{\nabla^{\mu}}
\newcommand{\nnu}{\nabla_{\nu}}

\newcommand{\fihc}{\frac{i}{\hbar c}}
\newcommand{\vpas}{4\pi\alpha_{{\rm S}}}
\newcommand{\vpias}{4\pi i\alpha_{{\rm S}}}

\newcommand{\fhcvpas}{\frac{\hbar c}{4\pi\alpha_{{\rm S}}}}
\newcommand{\fhcvpasezuinv}{\frac{\hbar c}{4\pi\alpha_{{\rm S}}{\rm e}^{-2u}}}
\newcommand{\fhcspias}{\frac{\hbar c}{16\pi\alpha_{{\rm S}}}}

\newcommand{\fhcapas}{\frac{\hbar c}{8\pi\alpha_{{\rm S}}}}
\newcommand{\tr}{{\rm tr}}

\newcommand{\vvr}{(\vec{r})}

\newcommand{\ihc}{i\hbar c}

\newcommand{\lk}{\left(}
\newcommand{\rk}{\right)}
\newcommand{\vn}{\vec{\nabla}}
\newcommand{\aM}{a_{{\rm M}}}
 
\newcommand{\MAmu}{{\mathcal A}_{\mu}}

\newcommand{\MAnu}{{\mathcal A}_{\nu}}

\newcommand{\exAmu}{{}^{({\rm ex})}\!A_{\mu}}

\newcommand{\exAOu}{{}^{({\rm ex})}\!A_{0}}
\newcommand{\exAou}{{}^{({\rm ex})}\!A_{0}}
\newcommand{\PAOu}{{}^{(p)}\!A_{0}}
\newcommand{\Amu}{A_{\mu}}
\newcommand{\Anu}{A_{\nu}}

\newcommand{\Aalomu}{A^{\alpha}{}_{\mu}}
\newcommand{\Aalonu}{A^{\alpha}{}_{\nu}}

\newcommand{\Abeomu}{A^{\beta}{}_{\mu}}

\newcommand{\Aeomu}{A^{1}{}_{\mu}}
\newcommand{\Azomu}{A^{2}{}_{\mu}}
\newcommand{\poAou}{{}^{({\rm p})}\!A_0}
\newcommand{\aoAou}{{}^{(a)}\!A_0}
\newcommand{\eoAou}{{}^{(1)}\!A_0}
\newcommand{\zoAou}{{}^{(2)}\!A_0}

\newcommand{\eAou}{{}^{(1)}\!A_0}
\newcommand{\zAou}{{}^{(2)}\!A_0}
\newcommand{\vAeu}{\vec{A}_1}
\newcommand{\vAzu}{\vec{A}_2}
\newcommand{\vA}{\vec{A}}

\newcommand{\Bmu}{B_{\mu}}

\newcommand{\Bsmu}{B^{*}_{\mu}}
\newcommand{\Bmo}{B^{\mu}}
\newcommand{\Bsmo}{B^{*\mu}}

\newcommand{\Bno}{B^{\nu}}

\newcommand{\Bsno}{B^{*\nu}}
\newcommand{\vB}{\vec{B}}
\newcommand{\vBs}{\vec{B}^{*}}

\newcommand{\rmd}{{\rm d}}
\newcommand{\MDmu}{{\mathcal D}_{\mu}}
\newcommand{\MDnu}{{\mathcal D}_{\nu}}
\newcommand{\MDmo}{{\mathcal D}^{\mu}}
\newcommand{\MDno}{{\mathcal D}^{\nu}}

\newcommand{\Dmu}{D_{\mu}}
\newcommand{\Dmo}{D^{\mu}}

\newcommand{\eu}{{\rm e}^u}

\newcommand{\ezu}{{\rm e}^{2u}}
\newcommand{\ezuinv}{{\rm e}^{-2u}}

\newcommand{\ETu}{E_{{\rm T}}}

\newcommand{\EeozoTuzu}{E^{(1,2)}_{{\rm T}}}

\newcommand{\EDu}{E_{{\rm D}}}
\newcommand{\ERu}{E_{{\rm R}}}
\newcommand{\ECu}{E_{{\rm C}}}
\newcommand{\Eesu}{E_{{\rm es}}}
\newcommand{\EeoRu}{E^{({\rm e})}_{{\rm R}}}
\newcommand{\EmoRu}{E^{({\rm m})}_{{\rm R}}}
\newcommand{\EhoCu}{E^{({\rm h})}_{{\rm C}}}
\newcommand{\EgoCu}{E^{({\rm g})}_{{\rm C}}}
\newcommand{\Eeoesu}{E^{({\rm e})}_{{\rm es}}}
\newcommand{\hEeoRu}{\hat{E}^{({\rm e})}_{{\rm R}}}
\newcommand{\tEeoRu}{\tilde{E}^{({\rm e})}_{{\rm R}}}
\newcommand{\hEmoRu}{\hat{E}^{({\rm m})}_{{\rm R}}}
\newcommand{\tEmoRu}{\tilde{E}^{({\rm m})}_{{\rm R}}}
\newcommand{\vE}{\vec{E}}
\newcommand{\MFmunu}{{\mathcal F}_{\mu\nu}}

\newcommand{\exFmunu}{{}^{({\rm ex})}\!F_{\mu \nu}}

\newcommand{\Fmunu}{F_{\mu \nu}}

\newcommand{\Falomunu}{F^{\alpha}{}_{\mu\nu}}
\newcommand{\Fbeomunu}{F^{\beta}{}_{\mu\nu}}
\newcommand{\Faomunu}{F^{a}{}_{\mu\nu}}

\newcommand{\Lfnu}{{}^{({\rm L})}\!f_{\nu}}

\newcommand{\Feomunu}{F^{1}{}_{\mu\nu}}
\newcommand{\Fzomunu}{F^{2}{}_{\mu\nu}}
\newcommand{\ggmu}{G_{\mu}}
\newcommand{\ggnu}{G_{\nu}}
\newcommand{\Gmunu}{G_{\mu\nu}}
\newcommand{\Gnumu}{G_{\nu\mu}}

\newcommand{\gmunu}{g_{\mu\nu}}

\newcommand{\Gsmunu}{G^{*}_{\mu \nu}}
\newcommand{\Gsnumu}{G^{*}_{\nu \mu}}

\newcommand{\gsu}{g_{*}}

\newcommand{\MHmu}{{\mathcal H}_{\mu}}

\newcommand{\MHmo}{{\mathcal H}^{\mu}}

\newcommand{\MHnu}{{\mathcal H}_{\nu}}

\newcommand{\MbHmu}{\overline{{\mathcal H}}_{\mu}}

\newcommand{\MbHnu}{{\overline{\mathcal H}}_{\nu}}

\newcommand{\vH}{\vec{H}}

\newcommand{\aoHjo}{{}^{(a)}\!H^j}
\newcommand{\MI}{{\mathcal I}}

\newcommand{\MJmu}{{\mathcal J}_{\mu}}

\newcommand{\MJnu}{{\mathcal J}_{\nu}}

\newcommand{\jmu}{j_{\mu}}

\newcommand{\jalomu}{j^{\alpha}_{\;\;\mu}}

\newcommand{\jaonu}{j^{a}{}_{\nu}}

\newcommand{\jaumu}{j_{a\mu}}

\newcommand{\jeomu}{j^{1}_{\;\;\mu}}

\newcommand{\jzomu}{j^{2}_{\;\;\mu}}

\newcommand{\jdomu}{j^{3}_{\;\;\mu}}
\newcommand{\jvomu}{j^{4}_{\;\;\mu}}
\newcommand{\jeonu}{j^{1}_{\;\;\nu}}
\newcommand{\jzonu}{j^{2}_{\;\;\nu}}

\newcommand{\jalumu}{j_{\alpha\mu}}

\newcommand{\jeumu}{j_{1\mu}}
\newcommand{\jzumu}{j_{2\mu}}

\newcommand{\jbeumu}{j_{\beta\mu}}

\newcommand{\jbomu}{j^{\beta}{}_{\mu}}

\newcommand{\jbeonu}{j^{\beta}{}_{\nu}}

\newcommand{\exjmu}{{}^{({\rm ex})}\!j_{\mu}}

\newcommand{\exjmo}{{}^{({\rm ex})}\!j^{\mu}}

\newcommand{\Kalubeu}{K_{\alpha\beta}}
\newcommand{\Kalobeo}{K^{\alpha\beta}}

\newcommand{\keumu}{k_{1\mu}}
\newcommand{\kzumu}{k_{2\mu}}
\newcommand{\keunu}{k_{1\nu}}
\newcommand{\kzunu}{k_{2\nu}}

\newcommand{\aokou}{{}^{(a)}\!k_0}

\newcommand{\akou}{{}^{(a)}\!k_0}
\newcommand{\ekou}{{}^{(1)}\!k_0}
\newcommand{\zkou}{{}^{(2)}\!k_0}
\newcommand{\Kpu}{K_{{\rm p}}}
\newcommand{\Ksu}{K_{{\rm s}}}

\newcommand{\LD}{L_{{\rm D}}}
\newcommand{\LG}{L_{{\rm G}}}
\newcommand{\LRST}{L_{{\rm RST}}}
\newcommand{\Lex}{L_{{\rm ex}}}
\newcommand{\aolmu}{{}^{(a)}\!l_{\mu}}

\newcommand{\aolnu}{{}^{(a)}\!l_{\nu}}
\newcommand{\eolnu}{{}^{(1)}\!l_{\nu}}
\newcommand{\zolnu}{{}^{(2)}\!l_{\nu}}
\newcommand{\aolou}{{}^{(a)}\!l_{0}}
\newcommand{\eolou}{{}^{(1)}\!l_{0}}
\newcommand{\zolou}{{}^{(2)}\!l_{0}}
\newcommand{\MM}{\mathcal{M}}

\renewcommand{\vr}{\vec{r}}
\newcommand{\Rpm}{R_{\pm}}

\newcommand{\Spm}{S_{\pm}}
\newcommand{\Tmunu}{T_{\mu\nu}}
\newcommand{\MTmunu}{\mathcal{T}_{\mu\nu}}

\newcommand{\esTmunu}{{}^{({\rm es})}\!T_{\mu\nu}}
\newcommand{\TTmunu}{{}^{({\rm T})}\!T_{\mu\nu}}
\newcommand{\TTouou}{{}^{({\rm T})}\!T_{00}}
\newcommand{\DTouou}{{}^{({\rm D})}\!T_{00}}
\newcommand{\GTouou}{{}^{({\rm G})}\!T_{00}}
\newcommand{\CTouou}{{}^{({\rm C})}\!T_{00}}
\newcommand{\esTouou}{{}^{({\rm es})}\!T_{00}}

\newcommand{\eTouou}{{}^{({\rm e})}\!T_{00}}
\newcommand{\gTouou}{{}^{({\rm g})}\!T_{00}}
\newcommand{\hTouou}{{}^{({\rm h})}\!T_{00}}
\newcommand{\DTmunu}{{}^{({\rm D})}\!T_{\mu\nu}}
\newcommand{\GTmunu}{{}^{({\rm G})}\!T_{\mu\nu}}
\newcommand{\CTmunu}{{}^{({\rm C})}\!T_{\mu\nu}}
\newcommand{\RTmunu}{{}^{({\rm R})}\!T_{\mu\nu}}
\newcommand{\eTmunu}{{}^{({\rm e})}\!T_{\mu\nu}}
\newcommand{\mTmunu}{{}^{({\rm m})}\!T_{\mu\nu}}
\newcommand{\ehTmunu}{{}^{({\rm e})}\!\hat{T}_{\mu\nu}}
\newcommand{\mhTmunu}{{}^{({\rm m})}\!\hat{T}_{\mu\nu}}
\newcommand{\etTmunu}{{}^{({\rm e})}\!\tilde{T}_{\mu\nu}}
\newcommand{\mtTmunu}{{}^{({\rm m})}\!\tilde{T}_{\mu\nu}}
\newcommand{\gTmunu}{{}^{({\rm g})}\!T_{\mu\nu}}
\newcommand{\hTmunu}{{}^{({\rm h})}\!T_{\mu\nu}}

\newcommand{\vX}{\vec{X}}

\newcommand{\vY}{\vec{Y}}

\newcommand{\zex}{z_{\rm{ex}}}
\newcommand{\hza}{\hat{z}_a}
\newcommand{\hze}{\hat{z}_1}
\newcommand{\hzz}{\hat{z}_2}
\newcommand{\aS}{\alpha_{{\rm S}}}

\newcommand{\Gmu}{\Gamma_{\mu}}
\newcommand{\Gmo}{\Gamma^{\mu}}
\newcommand{\Gnu}{\Gamma_{\nu}}

\newcommand{\gmu}{\gamma_{\mu}}

\newcommand{\zjlm}{\zeta{}^j{}_l^{,}{}^m}

\newcommand{\Smono}{\Sigma^{\mu\nu}}
\newcommand{\talu}{\tau_{\alpha}}
\newcommand{\tbu}{\tau_{\beta}}
\newcommand{\tgu}{\tau_{\gamma}}

\newcommand{\ualumu}{\upsilon_{\alpha\mu}}

\newcommand{\wmu}{\omega_{\mu}}

\begin{document}

\title{\LARGE{Relativistic Energy Levels \\ of Para-Helium}\\[6ex]}
\author{R. Gr\"abeldinger, P. Schust, M. Mattes and M. Sorg}
\affiliation{ II.\ Institut f\"ur Theoretische Physik der
Universit\"at Stuttgart\\Pfaffenwaldring 57\\ D-70550 Stuttgart\\ Germany\\
{\rm e-mail:} {\tt sorg@theo2.physik.uni-stuttgart.de}}
\begin{abstract}
\vfill
The practical usefulness of Relativistic Schr\"odinger Theory (RST) is tested by
calculating approximately the energy difference between the excited singlet state~$1s2s\,
{}^1S_0$ and the ground state~$1s^2\,{}^1S_0$ of the helium-like ions with arbitrary charge
number~$z_{\rm ex}\,(2\le z_{\rm ex}\le 100)$. The results are compared to the
corresponding predictions of other theoretical approaches in the literature and to the
experimental data. Since the exact solutions of the RST energy eigenvalue problem are
unknown, one has to resort to approximative methods. However the crudest approximation
(``spherically symmetric approximation'') yields relatively accurate results so that it
seems worth while to develop more powerful approximation techniques
\end{abstract}
\maketitle
\section{\label{s1}Introduction}
The well\bs known wave\bs particle duality \cite{b42} in quantum theory \cite{b61} enables one to interprete some of the quantum phenomena in terms of the particle picture \kl with the associated statistical interpretation\kr, whereas other phenomena of elementary matter are more conveniently understood in fluid\bs dynamic terms due to the wave picture, e.g. the Bose\bs Einstein condensates \cite{b40},\cite{b39}. Especially interesting are the fields of ``intersection'' where both pictures can be simultaneously applied, and then must be expected to predict the same numerical results for the outcome of the corresponding experiments. An example for this type is non\bs relativistic atomic physics where it can be shown that the fluid\bs dynamic approach \kl i.e. the density functional formalism \cite{b57,b58}\kr~is exactly equivalent to conventional quantum mechanics, being based on the use of wave functions in place of density distributions. \\
\indent The situation is somewhat different in \textit{relativistic} quantum mechanics where neither a well\bs working particle theory nor its fluid\bs dynamic counterpart for $N$\bs particle systems does exist \kl see the many critical comments on the Bethe\bs Salpeter equations, for instance in ref.s \cite{a23,b31}. \\
\indent The recently established {\bf R}elativistic {\bf S}chr\"odinger {\bf T}heory \kl RST\kr~\cite{a16,a42} is intended to fill just this theoretical gap where both basic concepts \kl i.e. wave functions and densities\kr~become united into one picture. From the mathematical point of view, this new theory is related rather to the fluid\bs dynamic than to the probabilistic view upon quantum theory. The reason for this is that RST takes the Whitney {\bf sum} of the single\bs particle bundles as the many\bs particle bundle, in contrast to conventional quantum mechanics which is based upon the tensor {\bf product} of the single\bs particle Hilbert spaces. Subsequently, we will present an example where the fluid\bs dynamic approach \kl i.e. RST\kr~is able to predict frequencies of atomic spectral lines with acceptable precision, albeit somewhat more inaccurate than conventional quantum mechanics does. It is important to place emphasis on the fact that we have to compare here our {\em approximative} RST results to those of other approximative methods, which have to be applied in the conventional domain because a consistent and complete {\em conventional} theory for relativistic $N$\bs particle systems does not exist. More concretely, we will apply RST in order to compute the energy difference between the excited singlet state $1s\;2s\;{}^{1}\!S_0$ and the ground state $1s^2\;{}^{1}\!S_0$ of the helium\bs like ions with arbitrary nuclear charge number $\zex$ \kl$2\leq\zex\leq100$\kr. Furthermore, we will compare our results to the analogous predictions obtained by the $1/Z$\bs expansion method \cite{a6} which are close to the analogous predictions of the all\bs order technique in relativistic many\bs body perturbation theory \kl MBPT\kr \cite{a5}. Since the latter two {\em approximation} techniques rely on the conventional {\bf tensor product} formalism, whereas RST utilises the {\bf Whitney sum}, the predictive potentiality of both formalisms can be directly opposed. And it is worth noting that RST's fluid\bs dynamic approach delivers predictions relatively close to the observational data; see tables I, II and figures 1,2. However, it must be stressed that the present comparison of numerical predictions for spectroscopic data between RST and the conventional approach is actually a competition between {\em approximative} methods. Therefore, the fact that the present RST results are somewhat more inaccurate than their conventional counterparts \cite{a6, a5} does not hint on an intrinsic deficiency of RST itself but merely points to the necessity of developing better approximation techniques within the framework of RST. Such an improvement may be conceived in a two\bs fold way, namely by
\begin{itemize}
\item[(i)] invention of more adequate trial functions for extremalising the RST action integral
\item[(ii)] solving the corresponding {\em two\bs dimensional} mass eigenvalue problem in the variables $r$, $\vartheta$ (in place of resorting to the spherically symmetric approximation).
\end{itemize}
\indent The results are elaborated through the following arrangement: \\
\indent For the convenience of the reader unfamiliar with RST, some of this new theory's fundamentals are collected into a brief survey, such as the {\em RST dynamics} in {\bf section \ref{s2}} and the {\em conservation laws} following from the RST dynamics in {\bf section \ref{s3}}. But of no lesser importance is the {\em RST kinematics}, such as the {\em fibre metric} $\Kalubeu$ for the Lie algebra bundle. Indeed, this object establishes an important link between the {\em RST currents} $\jalumu$ and the {\em Maxwell currents} $\jalomu$ emerging in the gauge field equations; see {\bf sections \ref{s4}} and {\bf \ref{s5}}. The existence of an {\em action principle} \kl{\bf section \ref{s6}}\kr~ is useful not only for the logical coherence of the theory and for the deduction of the conservation laws by means of Noether's theorem, but also for establishing approximation methods based upon variational techniques. The point here is that the RST eigenvalue problem for computing energy levels of bound $N$\bs particle systems is sufficiently complicate in order to prevent elaboration of analytic solutions. Therefore, we resort here to the {\em spherically symmetric approximation}. This method consists in adopting spherically symmetric ansatz functions for the wave amplitudes and substituting this {\em ``isotropic''} ansatz into the action principle in order to deduce the corresponding $SO(3)$\bs symmetric eigenvalue equations in {\bf section \ref{s7}} and {\bf \ref{s8}}. There exists an interesting point with this isotropic ansatz; namely, it is exact(!) for the states $ns^2\;{}^{1}\!S_0$, which share their symmetry with the ground\bs state $1s^2\;{}^{1}\!S_0$. The exact wave amplitudes of the more general solutions $n_1s\; n_2s\;{}^{1}\!S_0$ with different principal quantum numbers \kl$n_1\neq n_2$\kr~are not spherically symmetric but depend on both spherical polar coordinates $r$ and $\vartheta$. However, the isotropic ansatz with its exclusive dependence of the wave functions on the radial variable $r$ is tentatively considered as an acceptable approximation for those non\bs isotropic states \kl$n_1\neq n_2$\kr. \\
\indent The exactly isotropic states $ns^2\;{}^{1}\!S_0$ were already considered in great detail in some of the preceding papers and are therefore presented here only briefly in {\bf section \ref{s9}}, mainly in order to introduce the RST concept of self\bs interactions which is based on the fibre metric $\Kalubeu$. The latter object is fixed up to one free parameter $u$ \kl the {\em self\bs interaction parameter}\kr~whose determination has to be performed by adjusting the RST predictions to the corresponding experimental value for one selected physical situation (see the selection of the bismuth ionisation energy in \cite{a42,a43}).\\
\indent The spherically symmetric \kl and therefore approximative\kr~eigenvalue equations for the states $n_1 s\; n_2 s\;{}^{1}\!S_0$ with $n_1\neq n_2$ are presented in {\bf section \ref{s10}}, followed by the normalisation prescription for the corresponding solutions in {\bf section \ref{s10n}}. The occurrence of the self\bs interactions is much more manifest in the general case \kl$n_1\neq n_2$\kr~than it is in the exactly isotropic subset $ns^2\;{}^{1}\!S_0$ of section \ref{s9}. The latter subset is easily identified within the more general set $n_1s\;n_2s\;{}^{1}\!S_0$ by letting coincide the principal quantum numbers \kl$n_1=n_2$\kr, mass eigenvalues \kl$M_1=M_2$\kr, gauge potentials, etc. In this sense, the {\em exact} isotropic states $ns^2\;{}^{1}\!S_0$ are kinematically included in the larger but {\em approximative} set $n_1s\;n_2s\;{}^{1}\!S_0$. \\
\indent Finally, for the theory to be able to produce definite predictions, it needs an energy functional \kl$E_T$, say\kr~whose value upon the numerical solutions of the exact \kl$n_1=n_2$\kr~or approximative \kl$n_1\neq n_2$\kr~eigenvalue system yields the desired atomic energy levels. This {\em RST energy functional} is the subject of {\bf section \ref{s11}}; and the corresponding numerical results are displayed in {\bf section \ref{s12}}, see tables I, II and figures 1,2. The corresponding experimental values are taken from ref. \cite{m1}.
\section{\label{s2}RST Dynamics}
The RST dynamics subdivides into the matter dynamics, Hamiltonian dynamics, and gauge field dynamics. The matter dynamics consists of the {\bf R}elativistic {\bf S}chr\"odinger {\bf E}quation \kl RSE\kr for the wave function $\Psi$
\begin{equation}
\label{21}
i\hbar c\MDmu\Psi = \MHmu\Psi \;,
\end{equation}
or, resp., the {\bf R}elativistic von {\bf N}eumann {\bf E}quation \kl RNE\kr
\begin{equation}
\label{22}
\MDmu\MI = \fihc\left[\MI\MbHmu - \MHmu\MI\right] \;,
\end{equation}
if matter is to be described in terms of an {\em intensity matrix} $\MI$ rather than in terms of a wave function $\Psi$. Clearly, a {\em pure state} $\Psi$ can always be conceived as a special type of {\em mixture} $\MI$, namely, as the tensor product of $\Psi$ and its Hermitian conjugate $\overline{\Psi}$:
\begin{equation}
\label{23}
\MI \Rightarrow \Psi\otimes\overline{\Psi} \;.
\end{equation}
\indent The {\em Hamiltonian} $\MHmu$ is itself a dynamical object of the theory; and therefore it is to be determined by its own field equations, i.e. the {\em integrability condition}
\begin{equation}
\label{24}
\MDmu\MHnu - \MDnu\MHmu + \fihc\left[\MHmu,\MHnu\right] = i\hbar c\MFmunu \;,
\end{equation}
and the {\em conservation equation}
\begin{equation}
\label{25}
\MDmo\MHmu - \fihc\MHmo\MHmu = -i\hbar c\left\{\left(\frac{\MM c}{\hbar}\right)^2 + \Smono\MFmunu\right\} \;.
\end{equation}
The meaning of the integrability condition \rf{24} is to ensure the {\em bundle identities}, such as
\begin{align}
\label{26}
\left[\MDmu\MDnu - \MDnu\MDmu\right]\Psi &= \MFmunu\Psi \\
\label{27}
\left[\MDmu\MDnu - \MDnu\MDmu\right]\MI &= \left[\MFmunu,\MI\right] \;,
\end{align}
whereas the conservation equation \rf{25} guarantees the existence of conservation laws, for instance for the {\em total current} $\jmu$
\begin{equation}
\label{28}
\nmo\jmu \equiv 0 \;.
\end{equation}
Such a conservation law is indispensable for defining the particle number $N$ through
\begin{equation}
\label{29}
N = \int\limits_{(S)}\jmu{\rm d}S^{\mu} \;,
\end{equation}
where the choice of the hypersurface $(S)$ is arbitrary, just on behalf of the local law \rf{28}!
\section{\label{s3}Conservation Laws}
The matter densities, such as the total current $\jmu$ or the energy\bs momentum density $\DTmunu$, are quite generally defined by means of the intensity matrix $\MI$ and a corresponding operator \kl e.g. {\em velocity operator} $\Gmu$ or {\em energy\bs momentum operator} $\MTmunu$\kr~by the following trace prescription:
\begin{align}
\label{31}
\jmu &= \tr\lk\MI\cdot\Gmu\rk \\
\label{32}
{}^{({\rm D})}\!\Tmunu &= \tr\lk\MI\cdot\MTmunu\rk \;.
\end{align}
For the special case of $N$ identical Dirac particles, the {\em total velocity operator} $\Gmu$ is taken as the $N$\bs fold direct sum of the ordinary Dirac matrices $\gmu$
\begin{equation}
\label{33}
\Gmu = \underbrace{\gmu\oplus\gmu\oplus\ldots\oplus\gmu}_{N\mbox{\bs fold sum}} \;,
\end{equation}
and the energy\bs momentum operator $\MTmunu$ is assumed to be of the following form
\begin{equation}
\label{34}
\MTmunu = \frac{1}{4}\left\{\Gmu\MHnu + \MbHnu\Gmu + \Gnu\MHmu + \MbHmu\Gnu\right\} \;.
\end{equation}
\indent If the considered $N$\bs particle system is acted upon by an external field $\exFmunu$, the energy\bs momentum tensor of matter ${}^{({\rm D})}\!\Tmunu$ develops a non\bs trivial source which equals the {\em Lorentz force density} $\Lfnu$ \cite{a16,a42}
\begin{equation}
\label{35}
\nmo{}^{({\rm D})}\!\Tmunu = \Lfnu \;.
\end{equation}
Since this force density is composed of internal and external contributions, it does not vanish for vanishing external field strength $\exFmunu$. However, if one includes also the energy\bs momentum density of the gauge fields $\GTmunu$ and the external interaction density ${}^{({\rm es})}\!T_{\mu\nu}$ in order to obtain the {\em total energy\bs momentum density} $\TTmunu$ as
\begin{equation}
\label{35n}
{}^{({\rm T})}\!T_{\mu\nu} = {}^{({\rm D})}\!T_{\mu\nu} + {}^{({\rm G})}\!T_{\mu\nu} + {}^{({\rm es})}\!T_{\mu\nu} \;,
\end{equation}
then this total density has vanishing source whenever no external force field $\exFmunu$ due to an external source $\exjmu$ is present \kl i.e $\exFmunu = 0$, $\exjmu = 0$\kr:
\begin{equation}
\label{36}
\nmo\TTmunu = \hbar c\Fmunu\exjmo \Rightarrow 0 \;.
\end{equation}
\indent In contrast to this energy\bs momentum conservation, which does hold exclusively for closed systems only \kl$\exjmu = 0$\kr, the charge conservation law \rf{28} is valid in any case, whether the system is energetically closed or not. This can easily be demonstrated for the pure states \rf{23} when the following alternative version of the conservation equation \rf{25} is used \cite{a16,a42}
\begin{equation}
\label{37}
\Gmo\MHmu = \MM c^2 \;,
\end{equation}
which directly relates the Hamiltonian $\MHmu$ to the mass operator $\MM$. Indeed, multiplying both sides of the RSE \rf{21} by the total velocity operator $\Gmo$ from the left and observing \rf{37} yields just the Dirac equation for the $N$\bs particle wave function $\Psi$
\begin{equation}
\label{38}
i\hbar\Gmo\MDmu\Psi = \MM c\Psi \;.
\end{equation}
Furthermore, the total current $\jmu$ \rf{31} reads for the pure states $\Psi$ \rf{23}
\begin{equation}
\label{39}
\jmu = \overline{\Psi}\Gmu\Psi \;,
\end{equation}
so that the divergence of $\jmu$ vanishes just on account of the Dirac equation \rf{38} \kl and its Hermitian conjugate\kr~where the covariant constancy of the velocity operator $\Gmu$ has also been presumed
\begin{align}
\label{40}
\MDmu\Gnu &= 0 \\
\nonumber
\lk\MDmu\Gnu\right. &\doteqdot \left.\nmu\Gnu + \left[\MAmu,\Gnu\right]\rk \;.
\end{align}
\section{\label{s4}Currents}
In order to close the RST dynamics, one has to supply some field equation for the {\em bundle connection} $\MAmu$. This object enters the RST dynamics via the covariant derivative $\MDmu$ of the field objects, e.g. for the wave function $\Psi$ in the RSE \rf{21}
\begin{equation}
\label{41}
\MDmu\Psi \doteqdot \pmu\Psi + \MAmu\Psi \;,
\end{equation}
or similarly, for the Hamiltonian $\MHmu$ in the integrability condition \rf{24}
\begin{equation}
\label{42}
\MDmu\MHnu \doteqdot \nmu\MHnu + \left[\MAmu,\MHnu\right] \;.
\end{equation}
The choice for such a field equation for the ``gauge potential'' $\MAmu$ is not quite arbitrary because the desired equation has to be compatible with the already fixed RST dynamics. \\
\indent In order to exemplify this important point for a special RST object, consider the {\em gauge currents} $\jalumu$, \kl$\alpha = 1,\ldots,N^2$\kr
\begin{align}
\label{43}
\jalumu &= \tr\lk\MI\cdot\upsilon_{\alpha\mu}\rk \;,
\end{align}
with the {\em gauge velocity operators} $\upsilon_{\alpha\mu}$ being defined in terms of the gauge algebra generators $\talu$ through
\begin{equation}
\ualumu = \frac{i}{2}\left\{\talu,\Gmu\right\} \;.
\end{equation}
It can be shown that the RST dynamics implies the following source equations for these gauge currents $\jalumu$ \rf{43}:
\begin{equation}
\label{44}
\Dmo\jalumu \equiv 0 \;,
\end{equation}
with the covariant derivative $\Dmu$ being defined through
\begin{equation}
\label{45}
\Dmu j_{\alpha\nu} \doteqdot \nmu j_{\alpha\nu} - \omega^{\beta}{}_{\alpha\mu} j_{\beta\nu} \;.
\end{equation}
Here, the connection one\bs form $\wmu = \left\{\omega^{\beta}{}_{\alpha\mu}\right\}$ takes its values in the adjoint representation of the original connection $\MAmu$
\begin{equation}
\label{46}
\MAmu = A^{\alpha}{}_{\mu}\talu \;,
\end{equation}
i.e. one has
\begin{equation}
\label{47}
\omega^{\beta}{}_{\alpha\mu} = C^{\beta}{}_{\gamma\alpha}A^{\gamma}{}_{\mu} \;,
\end{equation}
with the $U(N)$ {\em structure constants} $C^{\beta}{}_{\gamma\alpha}$ being defined as usual:
\begin{equation}
\label{48}
\left[\talu,\tbu\right] = C^{\gamma}{}_{\alpha\beta}\tgu \;.
\end{equation}
\indent The source relation \rf{44} may also be recast into the more compact form
\begin{equation}
\label{49}
\MDmo\MJmu \equiv 0 \;,
\end{equation}
when the current operator $\MJmu$ is defined as follows:
\begin{equation}
\label{410}
\MJmu = i\jalomu\talu \;.
\end{equation}
Here, we have used the contravariant current components $\jalomu$ \kl''Maxwell currents''\kr~which are linked to their covariant counterparts $\jalumu$ \kl''RST currents''\kr~by means of some fibre metric $\Kalubeu$ ({\em compatibility tensor})
\begin{align}
\label{411}
\jalomu &= \Kalobeo\jbeumu \\
\label{412}
\jalumu &= \Kalubeu\jbomu \\
\nonumber
\lk\Kalubeu K^{\beta\gamma}\right. &= \left.\delta^{\gamma}{}_{\alpha}\rk \;,
\end{align}
which itself has to be covariantly constant:
\begin{equation}
\label{413}
\Dmu\Kalubeu \equiv 0 \;.
\end{equation}
Indeed, this covariant constancy of the fibre metric $\Kalubeu$ is necessary in order to ensure the source relation \rf{44} also in contravariant form
\begin{equation}
\label{413n}
\Dmo\jalomu = 0 \;,
\end{equation}
which, in turn, acts as the desired \textit{integrability condition} for the gauge field equations. Thus, the RST matter dynamics with its ``conservation law'' \rf{44} becomes {\em compatible} with the gauge field dynamics requiring the ``conservation law'' \rf{413n}.
\section{\label{s5}Gauge Field Equations}
After all these arrangements, it is now almost self\bs suggesting to select a field equation for the bundle connection $\MAmu$. For this purpose, first consider its curvature $\MFmunu$
\begin{equation}
\label{51}
\MFmunu \doteqdot \nmu\MAnu - \nnu\MAmu + \left[\MAmu,\MAnu\right] \;,
\end{equation}
and then postulate that this object obeys the \kl non\bs Abelian\kr Maxwell equations
\begin{align}
\label{52}
\MDmo\MFmunu &= -\vpias\MJnu \\
\nonumber
\Big(\aS\Big. &\doteqdot \Big.\frac{e^2}{\hbar c}\Big) \;.
\end{align}
In components, this gauge field equation reads
\begin{equation}
\label{53}
\Dmo\Fbeomunu = \vpas\jbeonu \;,
\end{equation}
provided the curvature $\MFmunu$ is decomposed into its components $\Falomunu$ \kl``field strengths''\kr~as usual; cf. \rf{46}
\begin{equation}
\label{54}
\MFmunu = \Falomunu\talu \;.
\end{equation}
Indeed, now one is easily convinced that the Maxwell equations \rf{52} are compatible with the preceding RST dynamics, and therefore also with their implication \rf{49}. Namely, it is just the generally valid bundle identity
\begin{equation}
\label{55}
\MDmo\MDno\MFmunu \equiv 0 \;,
\end{equation}
that has to be applied to the Maxwell equations \rf{52} in order to see that the continuity equation \rf{49} is actually satisfied whenever the Maxwell equations are obeyed.
\section{\label{s6}Action Principle}
The logical consistency of the coupled RST\bs Maxwell dynamics is supported by the fact that it can be deduced from an action principle \cite{a42}
\begin{equation}
\label{61}
\delta W_{{\rm RST}} = 0 \;,
\end{equation}
where the RST action integral is given as usual in terms of a Lagrangian $\LRST$ through
\begin{equation}
\label{62}
W_{{\rm RST}} = \int{\rm d}^4x\;\LRST\left[\Psi,\MAmu\right] \;.
\end{equation}
The preceding conservation laws may then be deduced from this action principle via the Noether theorems. According to the subdivision of the RST fields into a matter part $\Psi$ and a gauge field part $\MAmu$, the total Lagrangian $\LRST$ will be also composed of a matter contribution $\LD$ and a gauge field part $\LG$, together with a contribution $\Lex$ due to the external fields \kl e.g. Coulomb potential of a nucleus\kr:
\begin{equation}
\label{63}\LRST = \LD + \LG + \Lex \;.
\end{equation}
Here, the matter part $\LD$ reads in terms of the wave function $\Psi$
\begin{equation}
\label{64}
\LD = \frac{\ihc }{2}\left[\;\overline{\Psi}\Gmo\MDmu\Psi - \lk\MDmu\overline{\Psi}\rk\Gmo\Psi\right] - Mc^2\overline{\Psi}\Psi \;,
\end{equation}
the internal gauge field contribution $\LG$ is given in terms of the gauge potentials $\Aalomu$ as
\begin{align}
\label{65}
\LG &= \fhcspias K_{\beta\gamma}\Fbeomunu F^{\gamma\mu\nu} \\
\nonumber
 &\equiv \fhcspias \Fbeomunu F_{\beta}{}^{\mu\nu} \;,
\end{align}
and finally the external contribution $\Lex$ contains the external objects $\exAmu$, $\exFmunu$, $\exjmu$ in the following combination:
\begin{align}
\label{66}
\Lex &= \hbar c\left\{\exAmu\cdot j^{\mu} - \exjmu\cdot A^{\mu} - \frac{1}{8\pi\aS}\exFmunu F^{\mu\nu}\right\} \\
\nonumber
\Big( A_{\mu} &\doteqdot \Aeomu + \Azomu \;\;,\;\; \Fmunu \doteqdot \nmu\Anu - \nnu\Amu\Big)\;.
\end{align}
With this arrangement, the extremalisation of the action integral \rf{62} with respect to the wave function $\Psi$ yields the RST\bs Dirac equation \rf{38}; and similarly, the variation with respect to the gauge potential $\MAmu$ lets emerge the RST\bs Maxwell equation \rf{52}. \\
\indent Clearly, the scientific value of a new theory has to become evident by its successful application to practical problems. For this purpose, we choose here the energy difference $\Delta E_{1-2}$ of the para\bs helium states $1s2s\;{}^{1}\!S_0$ and $1s^2\;{}^{1}\!S_0$. The reason for this choice is that, on the one hand, for this quantity there do exist sufficient experimental data \cite{m1}, and on the other hand, the corresponding predictions of other theoretical approaches \cite{a6,a5} are also available. Thus, a comprehensive comparison of the predictive power of the various theoretical frameworks becomes feasible, both with inclusion or neglection of self\bs interactions.
\section{\label{s7}Stationary Bound States}
It should be a matter of course that the desired energy difference $\Delta E_{1-2}$ emerges also in RST by solving as usual the associated energy eigenvalue problem. The corresponding RST eigenvalue equations are to be deduced from the general RST dynamics by means of a stationary ansatz for the two\bs particle wave function $\Psi$
\begin{equation}
\label{71}
\Psi(x) = \lk\begin{array}{c} \psi_1(x) \\ \psi_2(x) \end{array}\rk \;.
\end{equation}
The ansatz for both Dirac spinors $\psi_a$, \kl$a=1,2$\kr~is as usual
\begin{equation}
\label{72}
\psi_a(\vec{r},t) = \exp\left[-i\frac{M_a c^2}{\hbar}t\right]\cdot\psi_a(\vec{r}) \;,
\end{equation}
with {\em mass eigenvalues} $M_a$ to be determined from the stationary form of the two\bs particle Dirac equation \rf{38}. For this purpose, the time\bs independent parts $\psi_a\vvr$ of the Dirac spinors \rf{72} are further decomposed into two\bs component Pauli spinors ${}^{(a)}\!\phi_{\pm}\vvr$ according to
\begin{align}
\label{73}
{}^{(1)}\!\phi_{+}\vvr &= R_{+}(r)\cdot\zeta^{\frac{1}{2}}{}^{,}_{0}{}^{\frac{1}{2}} \\
\label{74}
{}^{(1)}\!\phi_{-}\vvr &= -iR_{-}(r)\cdot\zeta{}^{\frac{1}{2}}{}^{,}_{1}{}^{\frac{1}{2}} \\
\label{75}
{}^{(2)}\!\phi_{+}\vvr &= S_{+}(r)\cdot\zeta{}^{\frac{1}{2}}{}^{,}_{0}{}^{-\frac{1}{2}} \\
\label{76}
{}^{(2)}\!\phi_{-}\vvr &= -iS_{-}(r)\cdot\zeta{}^{\frac{1}{2}}{}^{,}_{1}{}^{-\frac{1}{2}} \;.
\end{align}
Here, the usual spinor basis $\zjlm$ \cite{b27} has been used, obeying the angular\bs momentum relations
\begin{align}
\label{77}
\vec{J}^2\zjlm &= j\lk j +1\rk\hbar^2\zjlm \\
\label{78}
J_z\zjlm &= m\hbar\zjlm \\
\label{79}
\vec{L}^2\zjlm &= l\lk l+1\rk\hbar^2\zjlm \\
\label{710}
\vec{S}^2\zjlm &= s\lk s+1\rk\hbar^2\zjlm \;,
\end{align}
with the addition theorem for angular momentum
\begin{equation}
\label{711}
j = l \pm s \;.
\end{equation}
For electron spin $s=\frac{1}{2}$ this yields the configurations with $j=\frac{1}{2}$; $l=0,1$; $m=\pm\frac{1}{2}$ when restricted to singlet states ${}^1\!S_0$. It turns out that the presumed spherical symmetry \kl''isotropic ansatz''\kr~of the wave amplitudes $R_{\pm}(r)$, $S_{\pm}(r)$ is an exact symmetry for the states $ns^2\;{}^{1}S_0$. But for the general set $n_1s\;n_2s\;{}^{1}S_0$, this symmetry is adopted to be a useful approximation. \\
\indent In a similar way, the gauge fields $A^{\alpha}{}_{\mu}$ have to be specialised now to the stationary situation.
\section{\label{s8} Gauge Potentials}
First, the external potential $\exAmu = \lk\exAOu\vvr;-\vec{A}_{ex}\vvr\rk$ is assumed to be due to a fixed point\bs like nucleus of charge number $\zex$. Therefore, we put
\begin{equation}
\label{81}
\exAOu\vvr = \frac{\zex\aS}{r} \;,
\end{equation}
with no external magnetic field being present \kl i.e. $\vec{A}_{ex}\vvr \equiv 0$\kr. Next, the {\em electromagnetic potentials} $A^{a}{}_{\mu}$, $a=1,2$, cf. \rf{46}, are also assumed to be $SO(3)$ or $SO(2)$ symmetric, resp., and static:
\begin{align}
\label{82}
A^{a}{}_{\mu} &\rightarrow \left\{ {}^{(a)}\!A_0(r); -\vec{A}_a\vvr\right\}\ ,
\end{align}
where both electrostatic potentials ${}^{(a)}\!A_0(r)$ are additionally assumed to be identical for the special subset of isotropic states $ns^2\;{}^1\!S_0$:
\begin{equation}
\label{83}
{}^{(1)}\!A_0(r) \equiv {}^{(2)}\!A_0(r) \doteqdot {}^{(p)}\!A_0(r) \;.
\end{equation}
Since the spins point in opposite directions for para\bs configurations \kl$\leadsto$ singlet states\kr, the magnetostatic vector potentials are assumed to differ in sign for this special situation:
\begin{equation}
\label{84}
\vec{A}_1\vvr = -\vec{A}_2\vvr \doteqdot \vec{A}_p\vvr \;.
\end{equation}
\indent It is self\bs suggesting to evoke such a structure also for the electromagnetic currents $j_{a\mu}$ \kl$a=1,2$\kr, since the Maxwell equations \rf{53} link them to the electromagnetic potentials, i.e. we put
\begin{align}
\label{85}
\jeumu &= \overline{\psi}_2\gmu\psi_2 \doteqdot \kzumu = \left\{{}^{(2)}\!k_0(r);-\vec{k}_2\vvr\right\} \\
\label{86}
\jzumu &= \overline{\psi}_1\gmu\psi_1 \doteqdot \keumu = \left\{{}^{(1)}\!k_0(r);-\vec{k}_1\vvr\right\} \;.
\end{align}
Again, for the special subset of states $ns^2\;{}^{1}\!S_0$ the electrostatic {\em charge densities} ${}^{(a)}\!k_0(r)$ do coincide
\begin{equation}
\label{87}
{}^{(1)}\!k_0(r) \equiv {}^{(2)}\!k_0(r) \doteqdot {}^{(p)}\!k_0(r) \;,
\end{equation}
and the {\em three\bs currents} are antiparallel
\begin{equation}
\label{88}
\vec{k}_1\vvr = -\vec{k}_2\vvr \doteqdot \vec{k}_p\vvr \;.
\end{equation}
The simplest form of the three\bs current $\vec{k}_p\vvr$ and vector potential $\vec{A}_p\vvr$, to be used in the following for the singlet states $ns^2\;{}^1\!S_0$, reads in terms of spherical polar coordinates \kl$r,\vartheta,\varphi$\kr
\begin{align}
\label{89}
\vec{k}_p\vvr &= k_p(r)\sin \vartheta\cdot\vec{e}_{\varphi} \\
\label{810}
\vec{A}_p\vvr &= rA_p(r)\sin\vartheta\cdot\vec{e}_{\varphi} \;.
\end{align}
\indent Clearly, for the more general states $n_1s\;n_2s\;{}^1\!S_0$, the currents and potentials will have different strengths \kl$k_1\neq k_2$, $A_1\neq A_2$\kr:
\begin{align}
\label{811}
\vec{k}_1\vvr &= k_1(r)\sin\vartheta\vec{e}_{\varphi} \\
\label{812}
\vec{k}_2\vvr &= k_2(r)\sin\vartheta\vec{e}_{\varphi} \\
\label{813}
\vec{A}_1\vvr &= rA_1(r)\sin\vartheta\vec{e}_{\varphi} \\
\label{814}
\vec{A}_2\vvr &= rA_2(r)\sin\vartheta\vec{e}_{\varphi} \;,
\end{align}
but nevertheless, the scalars ${}^{(a)}\!k_0$, $k_a$, ${}^{(a)}\!A_0$, $A_a$ are assumed to be still of the spherically symmetric form. The latter assumption {\em spoils the exactness} of the solutions for $n_1\neq n_2$ which can hold exclusively for the special subset $n_1 = n_2$ \kl the exact but {\em non\bs isotropic} states $n_1s\;n_2s\;{}^1\!S_0$ require angular dependent ansatz functions and potentials: $R_{\pm}(r,\vartheta)$, $S_{+}(r,\vartheta)$, etc., see ref. \cite{u6}\kr. \\
\indent The exact $SO(3)$ symmetry of the $ns^2\:{}^1\!S_0$ configurations also suggests the following shape for the magnetic exchange potential $\vec{B}\vvr$ and associated current $\vec{h}\vvr$
\begin{align}
\label{815}
\vec{B}\vvr &= irB(r)\vec{W}_p\vvr \\
\label{816}
\vec{h}\vvr &= ih(r)\vec{W}^{*}_p\vvr \;,
\end{align}
where the complex\bs valued three\bs vector field $\vec{W}_p\vvr$ is given by
\begin{equation}
\label{817}
\vec{W}_p\vvr = -{\rm e}^{i\varphi}\kl\vec{e}_{\vartheta} + i\cos\vartheta\cdot\vec{e}_{\varphi}\kr \;.
\end{equation}
Here, the interesting point with the exchange four\bs potential $\Bmu=\left\{B_0,-\vec{B}\right\}$ is that the ground\bs state symmetry \kl i.e. ``isotropy''\kr~of the $ns^2\;{}^1\!S_0$ states demands the following identity of the magnetostatic potential $A_p(r)$ \rf{810} and magnetic exchange potential $B(r)$ \rf{815}:
\begin{equation}
\label{818}
B(r) \equiv A_p(r) \;.
\end{equation}
\indent For the exact non\bs isotropic states $n_1s\;n_2s\;{}^1\!S_0$ , the ansatz \rf{815} is retained, but now the assumption of spherical symmetry \kl i.e. putting $B(r,\vartheta)\Rightarrow B(r)$\kr spoils the exactness of the solution. Moreover, it is a general peculiarity of all singlet states ${}^1\!S_0$ that the {\em electric exchange potential} $B_0(r)$ is zero, which is a consequence of the vanishing of the exchange density $h_0$ as the time\bs component of the exchange current $h_{\mu}$
\begin{equation}
\label{819}
h_{\mu} \doteqdot \overline{\psi}_1\gmu\psi_2 = -\ezu j^4{}_{\mu} \equiv \ezu j^{3*}{}_{\mu} \;,
\end{equation}
with
\begin{equation}
\label{820}
h_0(r) = B_0(r) \equiv 0 \;.
\end{equation}
This peculiarity of the para\bs helium states ${}^{1}\!S_0$ implies that there is no exchange energy of the ``electric'' type which leaves the occurrence of the exchange energy to its ``magnetic'' counterpart being based upon the space parts $\vec{B}$, $\vec{h}$ of $B_{\mu}$ and $h_{\mu}$. Since these space parts $\vec{B}$, $\vec{h}$ are usually much smaller than the time components $B_0$, $h_0$,  it may appear that RST \kl and the Hartree\bs Fock approach as its non\bs relativistic limit\kr~produces too inaccurate predictions of the para\bs helium level system; see the discussion of this point in ref. \cite{u6}. The reason for this common deficiency of both the Hartree\bs Fock approach and RST is the following: this refers to the approximative assumption that the two\bs particle wave function could be approximated by only two (albeit appropriately chosen) one\bs particle wave functions (see the preceding RST ansatz being specified by equations \rf{71}\bs\rf{711}). Within the HF approach, the deficiency is superseded by resorting to the multiconfiguration Dirac\bs Fock method (MCDF) whose counterpart for curing the analogous RST deficiency will be presented in a separate paper.
\section{\label{s9}Mass Eigenvalue Equations for $\bf n_1=n_2$}
With the stationary form of all the RST fields being fixed, the next step is to substitute these expressions into the general two\bs particle Dirac equation \rf{38}. But only for the $ns^2\;{}^{1}\!S_0$ states, where both electrons share the same principle quantum number \kl$n_1=n_2\Rightarrow n$\kr, the present ``isotropic'' \kl i.e. spherically symmetric\kr~ansatz can be an exact solution. For this special class of solutions, the following identifications are self\bs evident:
\begin{align}
\label{91}
M_1 &= M_2 \doteqdot M^{\prime\prime} \\
\label{92}
R_{+}(r) &\equiv S_{+}(r) \\
\label{93}
R_{-}(r) &\equiv S_{-}(r) \\
\label{94}
{}^{(1)}\!k_0(r) &\equiv {}^{(2)}\!k_0(r) \doteqdot {}^{(p)}\!k_0(r) \\
\label{95}
k_1(r) &\equiv -k_2(r) \doteqdot k_p(r) \\
\label{96}
A_1(r) &\equiv -A_2(r) \equiv B(r) \;.
\end{align}
The corresponding eigenvalue system for the wave amplitudes $R_{\pm}(r)$ is exact and reads \cite{a16,a41}
\begin{align}
\label{97}
\frac{{\rm d}R_{+}(r)}{{\rm d}r} +\left[\exAOu(r) + \PAOu(r)\right]\cdot R_{-}(r) + 2rB(r)\cdot R_{+}(r) &= -\frac{M + M^{\prime\prime}}{\hbar}c\cdot R_{-}(r) \\
\label{98}
\frac{{\rm d}R_{-}(r)}{{\rm d}r} +\frac{2}{r}R_{-}(r) - \left[\exAOu(r) + \PAOu(r)\right]\cdot R_{+}(r) - 2rB(r)\cdot R_{-}(r) &= -\frac{M - M^{\prime\prime}}{\hbar}c\cdot R_{+}(r) \;.
\end{align}
Because the wave amplitudes $R_{\pm}(r)$ couple to the gauge fields $\PAOu(r)$ and $B(r)$, the eigenvalue system has to be closed by supplementing the Poisson equations for these gauge fields, which are of course to be deduced from the Maxwell equations \rf{52}:
\begin{align}
\label{99}
\lk\frac{{\rm d}^2}{{\rm d}r^2} + \frac{2}{r}\frac{{\rm d}}{{\rm d}r}\rk {}^{(p)}\!A_0(r) &= 4\pi\aS{}^{(p)}\!k_0(r) \\
\nonumber
 & \\
\label{910}
\lk\frac{{\rm d}^2}{{\rm d}r^2} + \frac{4}{r}\frac{{\rm d}}{{\rm d}r}\rk B(r) + 6B^2(r)\lk1-\frac{2}{3}r^2B(r)\rk &= 4\pi\aS\,\ezuinv\frac{k_p(r)}{r} \;.
\end{align}
Here, the gauge fields $\PAOu(r)$ and $B(r)$ couple back to the wave amplitudes via the charge and current densities ${}^{(p)}\!k_0(r)$ \rf{87} and $k_p(r)$ \rf{89} which read in terms of the wave amplitudes $R_{\pm}(r)$
\begin{align}
\label{911}
{}^{(p)}\!k_0(r) &= \frac{R^2_{+}(r) + R^2_{-}(r)}{4\pi}\\
\label{912}
k_p(r) &= \frac{R_{+}(r)\cdot R_{-}(r)}{2\pi} \;.
\end{align}
\indent A striking feature of the current mass eigenvalue system \rf{97}\bs\rf{910} is the occurrence of the self\bs interaction parameter $u$ in the last equation \rf{910}. Obviously, this is the only place where the self\bs interaction effect enters the eigenvalue system. It can be shown that this peculiarity is due to the fact that for the $ns^2\;{}^{1}S_0$ states both electrostatic charge densities coincide; see equation \rf{94}. More generally spoken, the self\bs interactions emerge in RST in connection with the covariantly constant fibre metric $\Kalubeu$ \rf{411}\bs\rf{413}. This objects converts {\em RST currents} $\jalumu$ to {\em Maxwell currents} $\jalomu$ which, in turn, enter the Maxwell equations \rf{53}. Thus, the fibre metric $\Kalubeu$ essentially determines the coupling of the field strengths $\Falomunu$ to the RST currents $\jalumu$ and thereby acquires the meaning of a {\em coupling matrix}. Its general shape for the currently considered two\bs particle system looks as follows \cite{a42}:
\begin{equation}
\label{913}
\left\{\Kalubeu\right\} = \lk\begin{array}{cccc} \eu\cdot\sinh u & -\eu\cdot\cosh u & 0 & 0 \\ -\eu\cdot\cosh u & \eu\cdot\sinh u & 0 & 0 \\ 0 & 0 & 0 & -\ezu \\ 0 & 0 & -\ezu & 0 \end{array}\rk \;.
\end{equation}
Evidently, this fibre metric $\Kalubeu$ owns one degree of freedom which plays the part of a ``renormalisation parameter'', because it continously changes the coupling strengths of gauge fields and currents. This is the reason why it emerges in connection with the original coupling constant $\aS=\frac{e^2}{\hbar c}$ on the right\bs hand side of equation \rf{910}. It is not subject to any constraints, and therefore, it can be used to let the RST predictions coincide with the observational data. The point here is that this matching of theoretical predictions and experimental data can be accomplished by fixing {\em one} value for $u$ for all the considered charge numbers $\zex$, see ref \cite{a43}. Observe also, that in the Poisson equation \rf{99} the {\em electrostatic} coupling constant $\aS$ is not renormalised by the self\bs interactions, in contrast to the {\em magnetic} case of equation \rf{910}: $\aS\Rightarrow\aS\ezuinv$. As a consequence of this peculiarity of the states $ns^2\;{}^1\!S_0$, the magnetic interactions are completely switched off for $u\to\infty$, leaving the field configuration to be of purely electrostatic type. This field configuration is already known from ref. \cite{u3} as the {\em electrostatic approximation}.
\section{\label{s10}Mass Eigenvalue Equations for $\bf n_1\neq n_2$}
It is necessary to stress that the eigenvalue system \rf{97}\bs\rf{910} for the $ns^2\;{}^{1}\!S_0$ states is {\em exact} and therefore can be deduced {\em either} from the RST variational principle \rf{61} by means of the isotropic ansatz for the wave functions $\psi_a$ and the gauge fields $\Aalomu$; {\em or} this system can be alternatively deduced by directly substituting the isotropic ansatz into the original field equations for matter \rf{38} and gauge fields \rf{52}. For a detailed study of the $ns^2\;{}^{1}\!S_0$ state see refs. \cite{a16,u6,a41,a43}. Generalising now the situation to the states $n_1s\;n_2s\;{}^{1}\!S_0$ with different principal quantum numbers $n_1$ and $n_2$, the isotropic ansatz can no longer be substituted into the RST field equations, because they do not admit those isotropic, well localised solutions for $n_1\neq n_2$ \kl albeit solutions, diverging at spatial infinity, may exist\kr. Nevertheless, the isotropic ansatz does not become useless for this more general situation, but it is merely devaluated to an {\em approximative} solution, whose quality may serve to estimate the influence of the neglected anisotropy. The corresponding approximate field equations are obtained by substituting the isotropic ansatz into the action principle \rf{61} and carrying out the variational procedure. This yields the following mass eigenvalue equations for the wave amplitudes $R_{\pm}$ and $S_{\pm}$ \cite{a43}:
\begin{align}
\label{101}
\frac{{\rm d}R_{+}}{{\rm d}r} + \left[\exAOu + {}^{(2)}\!A_0\right]\cdot R_{-} - \frac{2}{3}r\lk A_2 R_{+} - 2BS_{+}\rk &= -\frac{M_1 + M}{\hbar}c\cdot R_{-} \\
\label{102}
\frac{{\rm d}R_{-}}{{\rm d}r} + \frac{2}{r}\cdot R_{-} - \left[\exAOu + {}^{(2)}\!A_0\right]\cdot R_{+} + \frac{2}{3}r\lk A_2 R_{-} - 2BS_{-}\rk &= \frac{M_1 - M}{\hbar}c\cdot R_{+} \\
\label{103}
\frac{{\rm d}S_{+}}{{\rm d}r} + \left[\exAOu + {}^{(1)}\!A_0\right]\cdot S_{-} + \frac{2}{3}r\lk A_1 S_{+} + 2BR_{+}\rk &= -\frac{M_2 + M}{\hbar}c\cdot S_{-} \\
\label{104}
\frac{{\rm d}S_{-}}{{\rm d}r} + \frac{2}{r}\cdot S_{-} - \left[\exAOu + {}^{(1)}\!A_0\right]\cdot S_{+} - \frac{2}{3}r\lk A_1 S_{-} + 2BR_{-}\rk &= \frac{M_2 - M}{\hbar}c\cdot S_{+} \;.
\end{align}
\indent In a similar way, the Poisson equations for the electrostatic potentials ${}^{(a)}A_0(r)$ are found to be of the following form
\begin{align}
\nonumber
\lk\frac{{\rm d}^2}{{\rm d}r^2} + \frac{2}{r}\frac{{\rm d}}{{\rm d}r}\rk {}^{(1)}\!A_0(r) + \frac{8}{3}r^2 &B^2(r)\left[\frac{1}{\aM} - \lk{}^{(1)}\!A_0(r) - {}^{(2)}\!A_0(r)\rk\right] \\
\label{105}
 &= 4\pi\aS\cdot\ezuinv\left\{K_s(u)\cdot{}^{(2)}\!k_0(r) - K_p(u)\cdot{}^{(1)}\!k_0(r)\right\} \\
\nonumber
\lk\frac{{\rm d}^2}{{\rm d}r^2} + \frac{2}{r}\frac{{\rm d}}{{\rm d}r}\rk {}^{(2)}\!A_0(r) - \frac{8}{3}r^2 &B^2(r)\left[\frac{1}{\aM} - \lk{}^{(1)}\!A_0(r) - {}^{(2)}\!A_0(r)\rk\right] \\
\label{106}
 &= 4\pi\aS\cdot\ezuinv\left\{K_s(u)\cdot{}^{(1)}\!k_0(r) - K_p(u)\cdot{}^{(2)}\!k_0(r)\right\} \;.
\end{align}
Here, the {\em exchange length parameter} $\aM$ is defined as
\begin{equation}
\label{107}
\aM \doteqdot \frac{\hbar}{\lk M_1 - M_2\rk c} \;,
\end{equation}
the {\em pair coupling constant} $K_p(u)$ is given by
\begin{equation}
\label{108}
K_p(u) \doteqdot -\frac{\ezu + 1}{2} \;,
\end{equation}
and similarly, the {\em self\bs coupling constant} $K_s(u)$ by
\begin{equation}
\label{109}
K_s(u) \doteqdot \frac{\ezu - 1}{2} \;.
\end{equation}
\indent Concerning the magnetostatic potentials $A_a(r)$ \rf{813}\bs\rf{814} and the exchange potential $B(r)$ \rf{815}, the analogous variational procedure yields the following \kl generalised\kr Poisson equations
\begin{align}
\nonumber
\lk\frac{{\rm d}^2}{{\rm d}r^2} + \frac{4}{r}\frac{{\rm d}}{{\rm d}r}\rk A_1(r) + 6 &B^2(r)\left[1 - \frac{1}{3}r^2 \lk A_1(r) - A_2(r)\rk\right] \\
\label{1010}
 &= 4\pi\frac{\aS\ezuinv}{r}\left\{K_s(u)\cdot k_2(r) - K_p(u)\cdot k_1(r)\right\} \\
\nonumber
\lk\frac{{\rm d}^2}{{\rm d}r^2} + \frac{4}{r}\frac{{\rm d}}{{\rm d}r}\rk A_2(r) - 6 &B^2(r)\left[1 - \frac{1}{3}r^2 \lk A_1(r) - A_2(r)\rk\right] \\
\label{1011}
 &= 4\pi\frac{\aS\ezuinv}{r}\left\{K_s(u)\cdot k_1(r) - K_p(u)\cdot k_2(r)\right\}
\end{align}
\begin{align}
\nonumber
\lk\frac{{\rm d}^2}{{\rm d}r^2} + \frac{4}{r}\frac{{\rm d}}{{\rm d}r}\rk B(r) + B(r)\left\{\left[\frac{1}{\aM} - \lk{}^{(1)}\!A_0(r) - {}^{(2)}\!A_0(r)\rk\right]^2 + \right.\\
\label{1012}
\left. + 3\lk A_1(r) - A_2(r)\rk - r^2\left[2B^2(r) + \frac{1}{2}\lk A_1(r) - A_2(r)\rk^2\right]\right\} = 4\pi\aS\ezuinv\frac{h(r)}{r} \;,
\end{align}
with the current strengths $k_a(r)$ \rf{811}\bs\rf{812} and $h(r)$ \rf{816} being defined through
\begin{align}
\label{1013}
k_1(r) &= \frac{R_{+}(r)\cdot R_{-}(r)}{2\pi} \\
\label{1014}
k_2(r) &= -\frac{S_{+}(r)\cdot S_{-}(r)}{2\pi} \\
\label{1015}
h(r) &= \frac{R_{+}(r)\cdot S_{-}(r) + R_{-}(r)\cdot S_{+}(r)}{4\pi} \;.
\end{align}
\indent There is a nice consistency check for the more general situation with different quantum numbers $n_1\neq n_2$: the previous eigenvalue system \rf{97}\bs\rf{910} for the special subset of states $ns^2\;{}^{1}\!S_0$, being defined through the identifications \rf{91}\bs\rf{96}, must be recoverable from the present system \rf{101}\bs\rf{1012}. Indeed, some simple arguments show that this requirement is satisfied. In this sense, the more general system for $n_1\neq n_2$ appears to be the {\em unique} {\em isotropic} \kl and therefore approximative!\kr~generalisation of the special subcase with $n_1=n_2$. In view of this intimate relationship between the general and special cases, one may expect the general case yielding sufficiently realistic predictions for the $n_1s\;n_2s\;{}^{1}\!S_0$ energy levels despite its approximative character; see the numerical results below for verification of this supposition.
\section{\label{s10n}Normalisation Conditions}
The spectrum of desired solutions $R_{\pm}(r)$, $S_{\pm}(r)$ of the mass eigenvalue equations \rf{101}\bs\rf{104} is unique only if a normalisation condition upon these solutions is imposed. Such a condition is closely related to the asymptotic form of the electrostatic one\bs particle potentials ${}^{(a)}\!A_0\vvr$ \kl$a=1,2$\kr. Indeed, one expects that each of the two potentials ${}^{(a)}\!A_0\vvr$ is of the Coulomb form in the asymptotic region \kl$r\to\infty$\kr
\begin{equation}
\label{111}
{}^{(a)}\!A_0\vvr \Rightarrow -\frac{\aS}{r} \;,
\end{equation}
because each of the two electrons carries one negative charge unit. \\
\indent In order to see this asymptotic link between charge and potential more clearly, explicitely write down the first two equations of the RST\bs Maxwell system \rf{53} \kl i.e. for $\beta=1,2$\kr:
\begin{align}
\label{112}
\nmo\Feomunu &= \vpas\left\{\jeonu - \frac{i}{\vpas}\left[\Bmo\Gsmunu - \Bsmo\Gmunu\right]\right\} \\
\label{113}
\nmo\Fzomunu &= \vpas\left\{\jzonu + \frac{i}{\vpas}\left[\Bmo\Gsmunu - \Bsmo\Gmunu\right]\right\} \;.
\end{align}
Here the two\bs particle connection $\MAmu$ \rf{46} and its curvature $\MFmunu$ \rf{51} are decomposed with respect to the $U(2)$ generators $\left\{\tau_{\alpha}\right\}=\left\{\tau_1,\tau_2;\chi,\overline{\chi}\right\}$ as follows
\begin{align}
\label{114}
\MAmu &= -i\exAmu\cdot{\bf 1} + \Aeomu\tau_1 + \Azomu\tau_2 + \Bmu\chi - \Bsmu\overline{\chi} \\
\label{115}
\MFmunu &= -i\exFmunu\cdot{\bf 1} + \Feomunu\tau_1 + \Fzomunu\tau_2 + \Gmunu\chi - \Gsmunu\overline{\chi} \;.
\end{align}
Thus, defining the {\em entanglement vector} $\ggmu$ by
\begin{equation}
\label{116}
\ggmu \doteqdot \frac{i}{\vpas}\left[\Bno\Gsnumu - \Bsno\Gnumu\right] \;,
\end{equation}
both Maxwell equations \rf{112} and \rf{113} appear as
\begin{align}
\label{117}
\nmo\Feomunu &= -\vpas\eolnu \\
\label{118}
\nmo\Fzomunu &= -\vpas\zolnu \;,
\end{align}
provided the {\em effective currents} $\aolnu$ \kl$a=1,2$\kr~are defined in terms of the entanglement vector $\ggmu$ by
\begin{align}
\label{119}
\eolnu &= \ggnu - \jeonu \\
\label{1110}
\zolnu &= -\ggnu - \jzonu \;.
\end{align}
\indent The motivation for introducing such an arrangement is that these effective currents must have vanishing source
\begin{align}
\label{1111}
\nmo\,\aolmu &= 0 \;, \\
\nonumber
\left(a\right. &= \left.1,2\right)
\end{align}
in order that the Maxwell equations \rf{117}\bs\rf{118} be consistent. As a consequence, the desired normalisation conditions can be imposed upon the wave amplitudes in the following form:
\begin{equation}
\label{1112}
\int\limits_{(S)}\aolmu{\rm d}S^{\mu} = 1 \;,
\end{equation}
where the choice of the hypersurface $(S)$ in space\bs time is arbitrary just on  account of the source equations \rf{1111}. The validity of these important source equations can easily be checked by straightforward computation: First, write down the first two non\bs Abelian source equations \rf{413n} \kl i.e. for $\alpha=1,2$\kr
\begin{align}
\label{1113}
\nmo\jeomu &= i\left[\Bsmo\jdomu + \Bmo\jvomu\right] \\
\label{1114}
\nmo\jzomu &= -i\left[\Bsmo\jdomu + \Bmo\jvomu\right] \;,
\end{align}
and then form the divergence of the entanglement vector $\ggmu$ \rf{116} under use of the definition of the curvature components $\Falomunu$ \rf{54}
\begin{equation}
\label{1115}
\Falomunu = \nmu\Aalonu - \nnu\Aalomu + C^{\alpha}{}_{\beta\gamma}\Abeomu A^{\gamma}{}_{\nu} \;,
\end{equation}
which finally yields
\begin{equation}
\label{1116}
\nmo\ggmu = i\left[\Bsmo\jdomu + \Bmo\jvomu\right] \;.
\end{equation}
Thus, combining the last three equations \rf{1113}, \rf{1114} and \rf{1116} actually yields the desired source equations \rf{1111} for the effective currents $\aolmu$ \rf{119}\bs\rf{1110}. \\
\indent Clearly, for the stationary bound states considered here, the hypersurface $(S)$ of the normalisation integral \rf{1112} will be chosen as a time\bs slice \kl$t=const$\kr~which converts the general prescription \rf{1112} to an ordinary integral over three\bs space
\begin{align}
\label{1117}
\int & {\rm d}^3\vr\,{}^{(a)}l_0\vvr = 1 \\
\nonumber
& \left(a\right. = \left.1,2\right) \;.
\end{align}
Furthermore, according to the general relations \rf{411}, the Maxwellian currents $\jaonu$ may be expressed in terms of the RST currents $\jaumu$ \rf{85}\bs\rf{86} such that the effective currents $\aolmu$ \rf{119}\bs\rf{1110} finally reappear as
\begin{align}
\label{1118}
\eolnu &= \ggnu + \ezuinv\left\{\sinh u \cdot \kzunu + \cosh u \cdot \keunu\right\} \\
\label{1119}
\zolnu &= -\ggnu + \ezuinv\left\{\cosh u \cdot \kzunu + \sinh u \cdot \keunu\right\} \;.
\end{align}
Thus, defining the {\em normalisation parameters} $\hza$ \kl$a=1,2$\kr~as
\begin{equation}
\label{1120}
\hza \doteqdot \int{\rm d}^3\vec{r}\,\aokou\vvr \;,
\end{equation}
one finds that the normalisation conditions \rf{1117} can be transcribed to the following constraints for these normalisation parameters $\hza$:
\begin{align}
\label{1121}
\hze &= -\lk 1-g_{*}\rk\cdot K_p(u) - \lk 1+g_{*}\rk\cdot K_s(u) \\
\label{1122}
\hzz &= -\lk 1+g_{*}\rk\cdot K_p(u) - \lk 1-g_{*}\rk\cdot K_s(u) \;.
\end{align}
Here, the {\em exchange charge} $g_{*}$ is defined in terms of the time\bs component $G_0\vvr$ of the entanglement vector $\ggmu$ \rf{116} by
\begin{equation}
\label{1123}
g_{*} \doteqdot \int{\rm d}^3\vec{r}\,G_0\vvr \;,
\end{equation}
and the sum of both normalisation parameters $\hza$ \rf{1121}\bs\rf{1122} yields just the particle number \kl$N=2$\kr
\begin{equation}
\label{1124}
\hze + \hzz = 2 \;.
\end{equation}
\indent It is true, fixing the value of the normalisation parameters $\hza$ \rf{1120} by the constraints \rf{1121}\bs\rf{1122} really represents a normalisation condition upon the wave functions $\psi_a\vvr$, namely via the links \rf{85}\bs\rf{86} between charge densities $\aokou\vvr$ and wave functions $\psi_a\vvr$. But these conditions still depend on the gauge fields which enter the exchange charge $g_{*}$ \rf{1123}. Therefore, introducing the ``electric'' exchange field strength $\vX\vvr$ by
\begin{equation}
\label{1125}
\vX\vvr = \left\{X^j\vvr\right\} \doteqdot \left\{G_{0j}\vvr\right\} \;,
\end{equation}
the exchange charge reads in terms of the gauge fields
\begin{equation}
\label{1126}
g_{*} = \frac{i}{\vpas}\int{\rm d}^3\vvr\,\left[\vBs\vvr\cdot\vX\vvr - \vB\vvr\cdot\vX^{*}\vvr\right] \;.
\end{equation}
Furthermore, the vector field $\vX\vvr$ may be quite generally written in terms of the gauge potentials as \cite{a41}
\begin{equation}
\label{1127}
\vX\vvr = -\vec{\nabla}B_0\vvr + i\Delta_0\vvr\vB\vvr + iB_0\vvr\vec{\Delta}_{1,2}\vvr \;,
\end{equation}
where the scalar \kl$\Delta_0$\kr~and the vector \kl$\vec{\Delta}_{1,2}$\kr~potential differences are defined by \cite{a43}
\begin{align}
\label{1128}
\Delta_0\vvr &\doteqdot \frac{1}{\aM} - \left[\eoAou\vvr - \zoAou\vvr\right] \\
\label{1129}
\vec{\Delta}_{1,2}\vvr &\doteqdot \vAeu\vvr - \vAzu\vvr \;.
\end{align}
But fortunately, the ``electric'' exchange potential $B_0\vvr$ vanishes identically for the present spherically symmetric ansatz \rf{73}\bs\rf{76} so that the ``magnetic'' exchange potential $\vB\vvr$ becomes proportional to the ``electric'' exchange field strength $\vX\vvr$:
\begin{equation}
\label{1130}
\vX\vvr \Rightarrow i\Delta_0\vvr\cdot\vB\vvr \;.
\end{equation}
Using this simplification, together with the functional form \rf{815} for the exchange potential $\vB\vvr$, finally yields for the exchange charge $g_{*}$ \rf{1123}
\begin{equation}
\label{1131}
\gsu = -\frac{8}{3\aS}\int\limits_0^{\infty}{\rm d}r\,r^4\Delta_0(r)B^2(r) \;.
\end{equation}
\indent This result must be inserted into the normalisation conditions \rf{1121}\bs\rf{1122} and thereby renders the normalisation procedure an integral part of the eigenvalue problem itself so that both parts of the problem must be solved simultaneously. \kl For the subsequent numerical calculations, an iterative method will be applied where the normalisation \rf{1121}\bs\rf{1122} is repeatedly carried out on each iteration step\kr. For the exact subset of solutions \kl$n_1=n_2$\kr~of section \ref{s9}, the exchange charge $\gsu$ vanishes because the potential difference $\Delta_0\vvr$ \rf{1128} becomes zero; and thus, the normalisation conditions \rf{1121}\bs\rf{1122} are simplified to $\hza=1$. \\
\indent Finally, it remains to be shown that the postulated normalisation conditions \rf{1117} are consistent with the asymptotic Coulomb form \rf{111} of the electrostatic potentials $\aoAou\vvr$. For this purpose, rewrite the electrostatic Poisson equations \rf{105}\bs\rf{106} in terms of the effective charge densities $\aolou\vvr$ as
\begin{align}
\label{1132}
\Delta\eoAou\vvr &= \vpas\eolou\vvr \\
\label{1132n}
\Delta\zoAou\vvr &= \vpas\zolou\vvr \;.
\end{align}
\kl For the spherically symmetric approximation the effective charge densities $\aolou\vvr$ are here to be replaced by their average values over the solid angle $4\pi$\kr. Now, integrating these Poisson equations  \rf{1132}\bs\rf{1132n} over all three\bs space \kl$0\leq r<\infty$\kr~yields by means of Gau\ss' integral theorem \kl$a=1,2$\kr
\begin{equation}
\label{1133}
\lim_{r\to\infty}\frac{{\rm d}\aoAou\vvr}{{\rm d}r} = \frac{\aS}{r^2}\int{\rm d}^3\vr\,\aolou\vvr \;,
\end{equation}
so that the required asymptotic Coulomb form \rf{111} of the electrostatic potentials $\aoAou\vvr$ is guaranteed just by the normalisation conditions \rf{1117}! \kl For the topological character of the electromagnetic and exchange charges see ref \cite{a42}\kr.
\section{\label{s11}Energy Functional}
For the calculation of the energy levels \kl$E_T$, say\kr~of the two\bs electron atoms it is not sufficient to solve only the mass eigenvalue problem \rf{101}\bs\rf{1015}, because the mass eigenvalues $M_a$, $a=1,2$, do not immediately determine the energy levels $E_T$. The latter quantity is rather fixed by spatial integration of the total energy density $\TTouou$
\begin{equation}
\label{11n1}
E_T = \int\rmd^3\vr\;\TTouou\vvr \;,
\end{equation}
where, in a relativistic theory, the energy density $\TTouou\vvr$ is the time\bs time component of the corresponding energy\bs momentum density $\TTmunu$. This total density is found to consist essentially of three parts: the matter density $\DTmunu$, the gauge field density $\GTmunu$ and the interaction density $\esTmunu$ due to the external source $\exjmu$:
\begin{equation}
\label{11n2}
\TTmunu = \DTmunu + \GTmunu + \esTmunu \;.
\end{equation}
\indent Accordingly, the total energy $E_T$ \rf{11n1} is also built up by three contributions
\begin{equation}
\label{11n3}
E_T = E_D + E_G + E_{es} \;,
\end{equation}
with the self\bs evident identifications
\begin{align}
\label{11n4}
\EDu &= \int\rmd^3\vr\;\DTouou\vvr \\
\label{11n5}
\ETu &= \int\rmd^3\vr\;\GTouou\vvr \\
\label{11n6}
\Eesu &= \int\rmd^3\vr\;\esTouou\vvr \;.
\end{align}
Here, the gauge field density $\GTmunu$ turns out to consist of two subdensities $\RTmunu$ and $\CTmunu$ which are due to the electromagnetic modes $\Faomunu$ \kl$a=1,2$\kr and to the exchange modes $\Gmunu$, resp., cf. equation \rf{115}:
\begin{equation}
\label{11n7}
\GTmunu = \RTmunu - \CTmunu \;.
\end{equation}
Naturally, the electromagnetic part $\RTmunu$ appears as the sum of electric and magnetic contributions $\eTmunu$ and $\mTmunu$, i.e.
\begin{equation}
\label{11n8}
\RTmunu = \eTmunu + \mTmunu \;.
\end{equation}
A similar splitting does apply also to the exchange density $\CTmunu$
\begin{equation}
\label{11n9}
\CTmunu = \hTmunu + \gTmunu \;.
\end{equation}
This splitting of the exchange density $\CTmunu$ is completely analogous to the splitting of its electromagnetic counterpart \rf{11n8}; namely, the first part $\hTmunu$ is the energy\bs momentum density carried by the ``electric'' field strength $\vX\vvr$; and similarly, the second part $\gTmunu$ refers to the ``magnetic'' exchange field strength $\vY\vvr$. Furthermore, both electromagnetic contributions $\eTmunu$ and $\mTmunu$ are the sums of the mutual\bs interaction constituents $\ehTmunu$ and $\mhTmunu$ and of the self\bs interaction parts $\etTmunu$ and $\mtTmunu$:
\begin{align}
\label{11n10}
\eTmunu &= \ehTmunu + \etTmunu \\
\label{11n11}
\mTmunu &= \mhTmunu + \mtTmunu \;,
\end{align}
see ref \cite{a43}. Since each of these subdensities generates a contribution to the gauge field energy $\ETu$ via the recipes \rf{11n4}\bs\rf{11n6}, the total energy $\ETu$ \rf{11n3} finally appears as a sum of eight contributions
\begin{align}
\nonumber
\ETu &= \EDu + \lk\ERu - \ECu\rk + \Eeoesu \\
\label{11n12}
 &= \EDu + \lk\EeoRu + \EmoRu - \EhoCu - \EgoCu\rk + \Eeoesu \\
\nonumber
 &= \EDu + \lk\hEeoRu + \tEeoRu + \hEmoRu + \tEmoRu - \EhoCu - \EgoCu\rk + \Eeoesu \;.
\end{align}
\indent The following inspection of all of these energy contributions will yield further insight into the logical structure of RST.
\subsection{\label{s111}Matter Energy $\bf \EDu$}
First, consider the matter energy $\EDu$ \rf{11n4} which is the corrected sum of both mass\bs energies $M_a c^2$ \cite{a43}
\begin{equation}
\label{11n13}
\EDu = E_{{\rm D(1)}} + E_{{\rm D(2)}} = \lk\hze\cdot M_1 c^2 - \Delta E_{{\rm D(1)}}\rk + \lk\hzz\cdot M_2 c^2 - \Delta E_{{\rm D(2)}}\rk \;.
\end{equation}
The necessity of correction terms $\Delta E_{{\rm D}(a)}$, $a=1,2$, is immediately plausible because any mass eigenvalue $M_a$ \kl$a=1,2$\kr contains the interaction energies of each electron with the other one and with the external source; and therefore the simple sum of mass\bs energy equivalents $M_a c^2$ would embrace not only the sum of rest mass energy and kinetic energy as expected, see ref. \cite{a42}, but also the interaction  energies with the other particle and with the external source \kl i.e. nucleus\kr. This is the reason why those mutual interaction energies $\Delta E_{{\rm D}(a)}$ have to be subtracted from the individual mass eigenvalues $\hza\cdot M_a c^2$ in order to obtain the matter energy $\EDu$ \rf{11n13} exclusively as the sum of the kinetic energies of both particles and nothing else. \\
\indent Concerning the detailed structure of the subtracted terms $\Delta E_{{\rm D}(a)}$, one finds them to be a sum of three contributions refering to the external $({\rm es})$ and internal interactions, where the latter split up into the electrostatic $({\rm e})$ and exchange $({\rm h})$ types:
\begin{align}
\label{11n14}
\Delta E_{{\rm D}(a)} &= \hza\cdot M^{({\rm es})}_a c^2 +  \hza\cdot M^{({\rm e})}_a c^2 +  \hza\cdot M^{({\rm h})}_a c^2 \\
\nonumber
\Big(a &= 1,2\Big) \;.
\end{align}
The individual contributions are defined by \cite{u3}
\begin{align}
\label{11n15}
\hza\cdot M^{({\rm es})}_a c^2 &= -\hbar c \int\rmd^3\vr\;\exAou\vvr\cdot\akou\vvr \\
\label{11n16}
\hze\cdot M^{({\rm e})}_1 c^2 &= -\hbar c \int\rmd^3\vr\;\zAou\vvr\cdot\ekou\vvr \\
\label{11n17}
\hzz\cdot M^{({\rm e})}_2 c^2 &= -\hbar c \int\rmd^3\vr\;\eAou\vvr\cdot\zkou\vvr \\
\label{11n18}
\hze\cdot M^{({\rm h})}_1 c^2 &= \hzz\cdot M^{({\rm h})}_2 c^2 = \frac{\hbar c}{2}\int\rmd^3\vr\left\{B_0\vvr h_0\vvr + B^{*}_0\vvr h^{*}_0\vvr\right\} \;.
\end{align}
Clearly, the exchange contribution \rf{11n18} of electric type vanishes for the present para\bs states ${}^1\!S_0$ because here both fields $B_0\vvr$ and $h_0\vvr$ are zero, see the discussion below eq.\rf{820}.
\subsection{\label{s112}External Interaction Energy}
A further simplification of the total  functional $\ETu$ \rf{11n13} occurs in connection with the external contributions $M^{({\rm es})}_a c^2$ \rf{11n15}; and it is worthwhile to elaborate this point in detail because it elucidates the mechanism of electrostatic RST interactions. First, observe here that the sum of both external contributions \rf{11n13} brings into play the {\em total charge density} $j_0\vvr$:
\begin{align}
\label{11n19}
\sum\limits_{a=1}^{2}\hza\cdot M^{({\rm es})}_a c^2 &= -\hbar c\int\rmd^3\vr\;\exAou\vvr\cdot j_0\vvr \\
\label{11n20}
j_0\vvr &\doteqdot \ekou\vvr + \zkou\vvr \equiv j_{10}\vvr + j_{20}\vvr \equiv -\lk j^1{}_{0}\vvr + j^{2}{}_{0}\vvr \rk \;.
\end{align}
But the point with this total charge density $j_0\vvr$ is now that it generates the {\em total electrostatic potential} $A_0\vvr$, cf. \rf{66}
\begin{equation}
\label{11n21}
A_0\vvr \doteqdot \eAou\vvr + \zAou\vvr \;,
\end{equation}
via the {\em ordinary} \kl i.e. Abelian\kr~Poisson equation
\begin{equation}
\label{11n22}
\lk\frac{\rmd^2}{\rmd r^2} + \frac{2}{r}\frac{\rmd}{\rmd r}\rk A_0\vvr = \vpas j_0\vvr \;.
\end{equation}
Indeed, this claim is easily verified by simply adding up both {\em generalised} \kl i.e. non\bs Abelian\kr~Poisson equations \rf{105} and \rf{106}. Obviously, this interesting effect says that an {\em Abelian} structure arises for the {\em total} objects emerging from the sum of the {\em non\bs Abelian} electromagnetic objects! A further demonstration of this effect concerns the electric field strengths $\vec{E}_a\vvr\doteqdot\left\{{}^{(a)}\!F_{0j}\vvr\right\}$ being originally defined as
\begin{align}
\label{11n23}
\vE_1\vvr &= -\vn\,\eoAou\vvr - i\left[B_0\vvr\vBs\vvr - B_0^{*}\vvr\vB\vvr\right] \\
\label{11n24}
\vE_2\vvr &= -\vn\,\zoAou\vvr + i\left[B_0\vvr\vBs\vvr - B_0^{*}\vvr\vB\vvr\right] \;.
\end{align}
Thus, adding up both equations \rf{11n23} and \rf{11n24} yields again the ordinary \kl i.e. Abelian\kr~relationship between the total field strength $\vE\vvr$ and total potential $A_0\vvr$:
\begin{equation}
\label{11n25}
\vE\vvr \doteqdot \vE_1\vvr + \vE_2\vvr = -\vn A_0\vvr \;.
\end{equation}
\indent Presuming a similar relationship for the external objects
\begin{equation}
\label{11n26}
\vE_{{\rm ex}} = -\vn\exAou\vvr \;,
\end{equation}
enables one to reexpress the external interaction energy \rf{11n19} in terms of the electrostatic fields strengths $\vE\vvr$ \rf{11n25} and $\vE_{{\rm ex}}$ \rf{11n26}
\begin{equation}
\label{11n27}
\sum\limits_{a=1}^{2}\hza\cdot M^{({\rm es})}_a c^2 = -\frac{\hbar c}{\vpas}\int\rmd^3\vr\;\vE_{{\rm ex}}\cdot\vE\vvr \;.
\end{equation}
\indent This, however, is now a pleasant result because it yields a cancellation of the external gauge field energy $\Eesu$ \rf{11n6}. Indeed, the external interaction density $\esTmunu$
\begin{equation}
\label{11n28}
\esTmunu = \frac{\hbar c}{\vpas}\left\{{}^{({\rm ex})}\!F_{\mu\lambda}F_{\nu}{}^{\lambda} + {}^{({\rm ex})}\!F_{\nu\lambda}F_{\mu}{}^{\lambda} - \frac{1}{2}\gmunu{}^{({\rm ex})}\!F_{\sigma\lambda}F^{\sigma\lambda}\right\}
\end{equation}
is obviously a bilinear construction of the external field strengths $\exFmunu$ \rf{115} and the total field strength $\Fmunu$\kl$\doteqdot\Feomunu+\Fzomunu$\kr; and this arrangement transcribes to the electric part $\Eeoesu$ of the external interaction energy $\Eesu$ \rf{11n6} in the following way
\begin{align}
\label{11n29}
\Eeoesu &= \fhcvpas\int\rmd^3\vr\;\vE_{{\rm ex}}\cdot\vE\vvr \\
\nonumber
\Big(\vE\vvr &= \left\{E^j\vvr\right\} \doteqdot \left\{F_{0j}\vvr\right\}\;,\;\mbox{etc}\Big) \;.
\end{align}
Comparing this to the previous result \rf{11n27} verifies the claimed cancellation
\begin{equation}
\label{11n30}
-\sum\limits_{a=1}^{2}\hza\cdot M^{({\rm es})}_a c^2 + \Eeoesu = 0
\end{equation}
so that the sum of matter energy $\EDu$ \rf{11n13} and external interaction energy $\Eeoesu$ simplifies to
\begin{equation}
\label{11n31}
\EDu + \Eeoesu = \sum\limits_{a=1}^{2}\hza\cdot\lk M_a - M^{({\rm e})}_a\rk c^2 \;.
\end{equation}
Therefore, the final form of the total energy $\ETu$ \rf{11n12} looks as follows:
\begin{equation}
\label{11n32}
\ETu = \sum\limits_{a=1}^{2}\hza\cdot M_a c^2 + \lk\hEeoRu + \tEeoRu - \sum\limits_{a=1}^{2}\hza\cdot M^{({\rm e})}_a c^2\rk + \lk\hEmoRu + \tEmoRu\rk - \lk\EhoCu + \EgoCu\rk \;.
\end{equation}
This result states that the sum of mass energies is to be corrected by the electrostatic energy \kl first term in brackets\kr, by the magnetostatic energy \kl second term in brackets\kr, and by the exchange energy \kl third term in brackets\kr.
\subsection{\label{s113}Internal Electrostatic Interaction}
Naturally, there arises the question whether or not a similar cancellation like those for the {\em external} contribution in \rf{11n30} can occur also for the {\em internal} electrostatic gauge field energy, i.e. the first term in brackets on the right\bs hand side of equation \rf{11n32}. The self\bs energy $\tEeoRu$ and the mutual\bs interaction contribution $\hEeoRu$ of this term were previously identified as \cite{a43}
\begin{align}
\label{11n33}
\tEeoRu &= -\fhcapas K_{s}(u)\int\rmd^3\vr\;\left\{\vE_1^2\vvr + \vE_2^2\vvr\right\} \\
\label{11n34}
\hEeoRu &= -\fhcvpas K_{p}(u)\int\rmd^3\vr\;\vE_1\vvr\cdot\vE_2\vvr \;.
\end{align}
Consequently, integrating here by parts does not result in the same cancellation mechanism like that for the preceding external case \rf{11n30}. Instead, the following relationship between the electrostatic gauge field energy $\EeoRu$ and the mass\bs energy $M_a^{({\rm e})}c^2$ \rf{11n16}\bs\rf{11n17} holds:
\begin{align}
\label{11n35}
\EeoRu &\doteqdot \int\rmd^3\vr\;\eTouou\vvr = \hEeoRu + \tEeoRu = \frac{1}{2}\sum\limits_{a=1}^{2}\hza\cdot M_a^{({\rm e})} c^2 - N_{*}c^2 \;.
\end{align}
The mass\bs energies $M_a^{({\rm e})}c^2$ read in terms of the wave amplitudes $\Rpm$, $\Spm$
\begin{align}
\label{11n36}
\hze\cdot M_1^{({\rm e})} c^2 &= -\hbar c \int\rmd r\;r^2\left\{R_{+}^2 + R_{-}^2\right\}\zoAou(r) \\
\label{11n37}
\hzz\cdot M_2^{({\rm e})} c^2 &= -\hbar c \int\rmd r\;r^2\left\{S_{+}^2 + S_{-}^2\right\}\eoAou(r) \;,
\end{align}
and the additional energy content $N_{*}c^2$ due to the non\bs linear structure of the non\bs Abelian theory is given by
\begin{equation}
\label{11n38}
N_{*}c^2 = \frac{4}{3}\frac{\hbar c}{\aS\ezuinv}\int\rmd r\;r^4\Delta_0(r)\left[\frac{1}{\aM} - \Delta_0(r)\right]B^2(r) \;.
\end{equation}
\indent Evidently, the electrostatic mass\bs energy \rf{11n16}\bs\rf{11n17} of both electrons does not cancel against the gauge field energy $\EeoRu$ for the total energy $\ETu$ \rf{11n12} but, on the contrary, has still to be subtracted from the sum of mass eigenvalues together with the additional non\bs linear term $N_{*}c^2$. Thus, the total energy $\ETu$ \rf{11n32} appears now as
\begin{equation}
\label{11n39}
\ETu = \sum\limits_{a=1}^{2}\hza\cdot M_a c^2  - \lk\frac{1}{2}\sum\limits_{a=1}^{2}\hza\cdot M^{({\rm e})}_a c^2 + N_{*}c^2\rk + \lk\hEmoRu + \tEmoRu\rk - \lk\EhoCu - \EgoCu\rk \;.
\end{equation}
Clearly, this is again a plausible result because the sum of mass eigenvalues \kl first term on the right\bs hand side\kr~counts \textit{twice} the interelectronic interaction energy of the electrostatic type which therefore has to be subtracted \textit{once} \kl second term on the right\bs hand side\kr. Observe also, that this electrostatic interaction energy, being defined by the mean value of expressions \rf{11n16}\bs\rf{11n17}, contains no manifest self\bs coupling of the form $\aokou\cdot\aoAou$. But an indirect self\bs coupling does clearly exist because the $a$\bs th potential ${}^{(a)}A_0(r)$ is partly generated by the $b$\bs th wave function $\psi_b$ ($a\neq b$), see the electrostatic Poisson equations \rf{105}\bs\rf{106}.
\subsection{\label{s114}Internal Magnetostatic Interaction}
After the role of the electrostatic interactions has been clarified, one may turn attention to the somewhat different pattern of their magnetostatic counterparts. The mutual and self\bs interaction parts of this type of interactions were identified in terms of the magnetic fields $\vH_a\vvr$ as \cite{a43}
\begin{align}
\label{11n40}
\tEmoRu &= -\fhcapas K_{{\rm s}}(u)\int\rmd^3\vr\left\{\vH_1^2\vvr + \vH_2^2\vvr\right\} \\
\label{11n41}
\hEmoRu &= -\fhcvpas K_{{\rm p}}(u)\int\rmd^3\vr H_1\vvr\cdot\vH_2\vvr \;.
\end{align}
It is true, this looks quite similar to the electrostatic counterpart \rf{11n33}\bs\rf{11n34} but observe that for the magnetostatic interaction energy, in contrast to the electrostatic case, there is no need of removing it from the matter energy $\EDu$ \rf{11n13}. Another difference between both types of interactions is the non\bs Abelian construction of the magnetostatic fields $\vH_a\vvr$
\begin{equation}
\label{11n42}
\vH_a\vvr = \left\{\aoHjo\vvr\right\} \doteqdot \left\{\frac{1}{2}\epsilon^{jk}{}_{l}{}^{(a)}\!F_k{}^l\right\}
\end{equation}
by means of the potentials $\vA_a\vvr$, $\vB\vvr$; see ref \cite{a41}
\begin{align}
\label{11n43}
\vH_1\vvr &= \vn\times\vA_1\vvr - i\vB\vvr\times\vBs\vvr \\
\label{11n44}
\vH_2\vvr &= \vn\times\vA_2\vvr + i\vB\vvr\times\vBs\vvr \;.
\end{align}
This differs from the analogous situation with the electrostatic gradient fields $\vE_a\vvr$ \rf{11n23}\bs\rf{11n24}
\begin{align}
\label{11n45}
\vE_a\vvr &= -\vn\,\aoAou\vvr \\
\nonumber
\Big(a &= 1,2\Big) \;,
\end{align}
where the non\bs linear terms become zero on account of the vanishing ``electric'' exchange potential $B_0\vvr$! \\
\indent Clearly, this fact renders more complicated the mutual and self\bs interaction energies $\hEmoRu$ and $\tEmoRu$ of both particles
\begin{align}
\label{11n46}
\hEmoRu &= -\fhcvpas K_{{\rm p}}(u)\int\rmd^3\vr\vH_1\vvr\cdot\vH_2\vvr \\
\label{11n47}
\tEmoRu &= -\fhcapas K_{{\rm p}}(u)\int\rmd^3\vr\left\{\vH_1^2\vvr + \vH_2^2\vvr\right\} \;.
\end{align}
But on principle, the procedure is here the same as for the electrostatic case, i.e. one inserts the functional form of the magnetic fields $\vH_a\vvr$ \kl due to the isotropic ansatz\kr
\begin{align}
\label{11n48}
\vH_1\vvr &= 2\cos\vartheta\left[A_1(r)-r^2 B^2(r)\right]\cdot\vec{e}_r - \frac{\sin\theta}{r}\frac{\rmd}{\rmd r}\lk r^2 A_1(r)\rk\cdot\vec{e}_{\vartheta} \\
\label{11n49}
\vH_2\vvr &= 2\cos\vartheta\left[A_2(r)+r^2 B^2(r)\right]\cdot\vec{e}_r - \frac{\sin\theta}{r}\frac{\rmd}{\rmd r}\lk r^2 A_2(r)\rk\cdot\vec{e}_{\vartheta} \;,
\end{align}
and then one finds the magnetostatic energy functionals \rf{11n46} and \rf{11n47} by partially integrating and use of the magnetostatic Poisson equations \rf{1010} and \rf{1011} to be of the following form:
\begin{align}
\nonumber
\hEmoRu ={}& \frac{2}{3}\hbar c\Kpu\ezuinv\bigg[\int\rmd r\,r^3\left\{A^{\prime}_2\cdot R_{+}R_{-} - A^{\prime}_1\cdot S_{+}S_{-}\right\} + \frac{1}{\aS\ezuinv}\int\rmd r\,r^4 B^2\Big\{\Delta_{1,2}\lk1-r^2\Delta_{1,2}\rk \\
\label{11n50}
 &+ 2r^2B^2\Big\}\bigg] \\
\nonumber
\tEmoRu ={}& \frac{2}{3}\hbar c\Ksu\ezuinv\bigg[\int\rmd r\,r^3\left\{A^{\prime}_1\cdot R_{+}R_{-} - A^{\prime}_2\cdot S_{+}S_{-}\right\} - \frac{1}{\aS\ezuinv}\int\rmd r\,r^4 B^2\Big\{\Delta_{1,2}\lk1-r^2\Delta_{1,2}\rk \\
\label{11n51}
 &+ 2r^2B^2\Big\}\bigg] \;.
\end{align}
Despite the formal similarities of these magnetostatic energy contributions $\hEmoRu$, $\tEmoRu$ \rf{11n46} and \rf{11n47} with their electromagnetic counterparts $\hEeoRu$, $\tEeoRu$ \rf{11n33} and \rf{11n34} there is an important difference which refers to the occurrence of {\em direct self\bs interactions} for the magnetostatic case; see the remarks below equation \rf{11n39} for the corresponding electrostatic situation. The point with the magnetic case refers here to the occurrence of modified magnetostatic potentials $A^{\prime}_{a}(r)$ which are linked to the original $A_a(r)$ \rf{813}-\rf{814} in the following way:
\begin{align}
\label{11n52}
A^{\prime}_1(r) &= \Ksu(u)\cdot A_2(r) - \Kpu(u)\cdot A_1(r) \\
\label{11n53}
A^{\prime}_2(r) &= \Ksu(u)\cdot A_1(r) - \Kpu(u)\cdot A_2(r) \;.
\end{align}
This link clearly reveals the existence of direct self\bs interactions of the magnetic type since, e.g., the {\em second} modified potential $A^{\prime}_2$ in \rf{11n50}, on the one hand, combines with the wave amplitudes $\Rpm$ of the first particle, but on the other hand, it contains the {\em first} potential $A_1(r)$ which itself is generated just by the first particle \kl$a=1$\kr itself. In this way, the first particle acts back to itself:
\begin{equation}
\label{11n54}
\Rpm\Rightarrow A_1(r)\Rightarrow A^{\prime}_2(r)\Rightarrow A^{\prime}_2\cdot R_{+}R_{-}\Rightarrow\hEmoRu \;.
\end{equation}
This reasoning holds analogously for the first magnetostatic mode $A^{\prime}_1$. But clearly, an even more manifest self\bs interaction is established by the proper self-energy $\tilde{E}_R^{(m)}$ \rf{11n51} where the first (second) magnetostatic potential $A^{\prime}_1$ ($A^{\prime}_2$) directly combines with the first (second) current intensity $k_1$ ($k_2$), cf. eq.s \rf{1013}-\rf{1014}.
\subsection{\label{s115}Exchange Interactions}
The last contribution to the total energy is the exchange energy $\ECu$, which is the energy content of the exchange modes $\Gmunu$ as specified by $\CTmunu$ \rf{11n9}, i.e.
\begin{equation}
\label{11n55}
\ECu = \int\rmd^3\vr\,\CTouou(\vec{r}) \;.
\end{equation}
Since the energy\bs momentum density $\CTmunu$ \rf{11n9} of the exchange modes splits up into an ``electric'' part $\hTmunu$ and a magnetic part $\gTmunu$, this splitting transcribes to the exchange energy $\ECu$ \rf{11n55}:
\begin{align}
\label{11n56}
\ECu &= \EhoCu + \EgoCu \\
\label{11n57}
\EhoCu &\doteqdot \int\rmd^3\vr\,\hTouou\vvr \\
\label{11n58}
\EgoCu &\doteqdot \int\rmd^3\vr\,\gTouou\vvr \;.
\end{align}
Concerning the ``electric'' part $\EhoCu$, it is true that both the exchange density $h_0\vvr$ and the associated scalar exchange potential $B_0\vvr$ do vanish identically for the para\bs helium states ${}^1\!S_0$. But nevertheless, the exchange field strength $\vX\vvr$ is non\bs trivial, cf. \rf{1130}. Therefore, the corresponding exchange energy $\EhoCu$ of the ``electric'' type is non\bs zero and reads, quite generally, in terms of the exchange field strength $\vX\vvr$ \cite{a43}
\begin{equation}
\label{11n59}  
\EhoCu = \fhcvpasezuinv\int\rmd^3\vr\,\vX^{*}\vvr\cdot\vX\vvr \;,
\end{equation}
i.e. by use of the exchange field strength $\vX\vvr$ \rf{1130}
\begin{equation}
\label{11n60}
\EhoCu = \frac{4}{3}\frac{\hbar c}{\aS\ezuinv}\int\rmd r\,r^4\Delta_{0}^2(r)B^2(r) \;.
\end{equation}
\indent The final contribution to the total energy $\ETu$ \rf{11n39} is the exchange energy of ``magnetic'' type \cite{a43}
\begin{align}
\label{11n61}
\EgoCu & = \fhcvpasezuinv\int\rmd^3\vr\,\vY^{*}\vvr\cdot\vY\vvr \\
\nonumber
\bigg( \vY  & = \Big\{ Y^j \Big\} \doteqdot \Big\{ \frac{1}{2}\epsilon^{jk}{}_{l}G_k{}^l \Big\} \bigg) \;.
\end{align}
Decomposing here the ''magnetic'' exchange field strength $\vY\vvr$ with respect to spherical polar coordinates as usual
\begin{equation}
\label{11n62}
\vY\vvr = Y_r\vec{e}_r + Y_{\vartheta}\vec{e}_{\vartheta} + Y_{\varphi}\vec{e}_{\varphi} \;,
\end{equation}
with components
\begin{align}
\label{11n63}
Y_r &= -2\sin\vartheta\;{\rm e}^{i\varphi}B(r)\left[1-\frac{1}{2}r^2\Delta_{1,2}(r)\right] \\
\label{11n64}
Y_{\vartheta} &= -\cos\vartheta\;{\rm e}^{i\varphi}\frac{1}{r}\frac{\rmd}{\rmd r}\left[r^2B(r)\right] \\
\label{11n65}
Y_{\varphi} &= -i{\rm e}^{i\varphi}\frac{1}{r}\frac{\rmd}{\rmd r}\left[r^2 B(r)\right] \;,
\end{align}
ultimately yields by means of integrating by parts and use of the exchange Poisson equation \rf{1012}
\begin{align}
\nonumber
\EgoCu ={} \frac{4}{3}\hbar c\bigg\{\int\rmd r\,r^3B(r)\left[R_{+}\cdot S_{-} + R_{-}\cdot S_{+}\right] - \frac{1}{\aS\ezuinv}\int\rmd r\,r^4B^2(r) & \bigg[\Delta_0^2(r) + \Delta_{1,2}(r)\\
\label{11n66}
 & - 2r^2B^2(r)\Big]\bigg\} \;.
\end{align}
\indent With this result, the total energy $\ETu$ is completely fixed in terms of the wave amplitudes $\Rpm(r)$, $S_{\pm}(r)$ and the gauge potentials $\aoAou(r)$, $A_a(r)$ and $B(r)$. Its value upon the solutions of the energy eigenvalue problem, which consists of the mass eigenvalue equations and the Poisson equations, will yield the desired energy levels of the ground state $E_T^{(1,1)}$ and the first excited state $\EeozoTuzu$ in the next section.
\subsection{\label{s116}Special Case: $\bf ns^2\;{}^{1}\!S_0$}
As a check for the more general results concerning the states $n_1s\;n_2s\;{}^{1}\!S_0$, return to the subset of exactly isotropic states $ns^2\;{}^{1}\!S_0$ for which the total energy $\ETu$ was previously determined as \cite{a43}:
\begin{equation}
\label{11n67}
\ETu = 2M^{\prime\prime}c^2 - \EeoRu + 3\EmoRu \;,
\end{equation}
with the electrostatic energy $\EeoRu$ being given by
\begin{equation}
\label{11n68}
\EeoRu = -\hbar c\int\rmd r\,r^2\,\left[R^2_{+}(r) + R^2_{-}(r)\right]\poAou(r) \;,
\end{equation}
and its magnetic counterpart $\EmoRu$ by
\begin{equation}
\label{11n69}
\EmoRu = \frac{4}{3}\hbar c \left\{\int\limits_{0}^{\infty}\rmd r\,r^3B(r)R_{+}(r)R_{-}(r) - \frac{1}{\aS\ezuinv}\int\limits_{0}^{\infty}\rmd r\,r^4 B^3(r)\left[1-r^2B(r)\right]\right\} \;.
\end{equation}
Observe the prefactor of three for the magnetostatic contribution $\EmoRu$ \rf{11n67} to the total energy $\ETu$. It was shown in the preceeding articles \cite{a16,a41} that this threefold weight of $\EmoRu$ relative to $\EeoRu$ is due to the exact isotropy of the states $ns^2\;{}^{1}\!S_0$, where each of the three coordinate axis contributes the same amount to the spin\bs spin interaction energy. \\
\indent The special form \rf{11n67} of the energy functional $\ETu$ is immediately obtainable from the more general result \rf{11n39} by use of the former identifications \rf{91}\bs\rf{96}. They put the exchange charge $\gsu$ \rf{1126} to zero because the ``electric'' exchange field strength $\vX\vvr$ vanishes  as does the electrostatic potential difference $\Delta_0(r)$ \rf{1128}. As a consequence, both normalisation parameters $\hza$ \rf{1121}\bs\rf{1122} adopt unity value, $\hza\Rightarrow 1$. In turn, this result reduces the general sum of mass eigenvalues occuring in the energy functional $\ETu$ \rf{11n39} to the special form \rf{11n67}:
\begin{equation}
\label{11n70}
\sum\limits_{a=1}^{2}\hza\cdot M_ac^2 \Rightarrow 2M^{\prime\prime}c^2 \;.
\end{equation}
\indent Next, pay attention to the vanishing of the non\bs linear term $N_{*}c^2$ \rf{11n38} which is due to the vanishing electrostatic potential difference $\Delta_0(r)$ \rf{1128}:
\begin{equation}
\label{11n71}
N_{*}c^2 \Rightarrow 0\;.
\end{equation}
Thus, the remainder of the electrostatic interactions equals half the sum of mass equivalents $
M_a^{({\rm e})}$ \rf{11n16}\bs\rf{11n17} as being specified by the results \rf{11n36}\bs\rf{11n37}. By the identifications \rf{91}\bs\rf{96} the latter two equations adopt the following special form
\begin{equation}
\label{11n72}
\frac{1}{2}\sum\limits_{a=1}^{2}\hza\cdot M_a^{({\rm e})}c^2 \Rightarrow -\hbar c\int\rmd r\,r^2\,\poAou(r)\left\{R^2_{+} + R^2_{-}\right\} \;.
\end{equation}
This is just the required result for the electromagnetic interaction energy $\EeoRu$ \rf{11n68} due to the states $ns^2\;{}^1\!S_0$. \\
\indent Next, consider the general magnetostatic contributions $\hEmoRu$ \rf{11n50} and $\tEmoRu$ \rf{11n51}. Again, applying to these expressions the identifications \rf{91}\bs\rf{96} simplifies them to
\begin{align}
\label{11n73}
\hEmoRu \Rightarrow -\frac{4}{3}\hbar c\Kpu(u)\ezuinv\left\{\int\rmd r\,r^3 R_{+}(r)R_{-}(r)B(r) - \frac{1}{\aS\ezuinv}\int\rmd r\,r^4B^3(r)\left[1-r^2B(r)\right]\right\} \\
\label{11n74}
\tEmoRu \Rightarrow -\frac{4}{3}\hbar c\Ksu(u)\ezuinv\left\{\int\rmd r\,r^3 R_{+}(r)R_{-}(r)B(r) - \frac{1}{\aS\ezuinv}\int\rmd r\,r^4B^3(r)\left[1-r^2B(r)\right]\right\}\;.
\end{align}
Therefore, for the exactly isotropic states $ns^2\;{}^1\!S_0$, one adds up both equations \rf{11n73} and \rf{11n74} and thus finds the required special form \rf{11n69}. \\
\indent To conclude matters, call your attention to $\EhoCu$ \rf{11n60} and $\EgoCu$ \rf{11n66}, the yet unconsidered exchange energy contributions of ``electric'' and ``magnetic'' type to the total energy $\ETu$ \rf{11n39}. Clearly, the ``electric'' contribution $\EhoCu$ vanishes because of the vanishing electrostatic potential difference $\Delta_0(r)$ \rf{1128}:
\begin{equation}
\label{11n75}
\EhoCu \Rightarrow 0 \;.
\end{equation}
Up to sign, the magnetic contribution $\EgoCu$ \rf{11n66} adopts the double value of the magnetostatic contribution $\EmoRu$ \rf{11n69}:
\begin{equation}
\label{11n76}
\EgoCu \Rightarrow -2\EmoRu \;.
\end{equation}
So, in summary, the special form \rf{11n67} of the energy functional $\ETu$ for the exact isotropic states does indeed emerge from the general shape \rf{11n39}, namely by substituting therein the special results \rf{11n70}\bs\rf{11n76} referring to the states $ns^2\;{}^1S_0$.
\section{\label{s12}Numerical Results}
For a comparison of the present RST results to the experimental data and to the corresponding predictions of other theoretical approaches, observe that the literature mostly presents {\em ionisation energies} $\kl J^{({\rm 1,n})}_{{({\bf 2})}}$, say$\kr$~which refer to the states of the helium\bs like ions, where one of both electrons occupies the groundstate $\lk 1s\rk$ and the other one is in an excited state $\lk ns\rk$.  \\
Hence, comparing the various theoretical predictions for the chosen energy difference
\begin{equation}
\label{121}
\Delta_{{\rm theo}}E_{1-2} \doteqdot \left.E_{{\rm theo}}\right|_{1s\;2s\;{}^{1}\!S_0} - \left.E_{{\rm theo}}\right|_{1s^2\;{}^{1}\!S_0}
\end{equation}
among each other and all of them to the experimental values $\Delta_{{\rm exp}}E_{1-2}$, has to be preceded by transcribing the ionisation energies $J^{({\rm 1,2})}_{{\rm theo(2)}}$ to the desired energy difference \rf{121}. This can be easily done by simply observing that the concept of ``ionisation energy'' refers to the energy difference between a {\em final} state and an \textit{initial} state, where in the final state one of both electrons is free, having energy $Mc^2$, and the other one occupies a bound state with energy $E_{{\rm theo({\bf 1})}}^{(1)}$. On the other hand, in the initial state both electrons do occupy a two\bs particle state of energy $E_{{\rm theo({\bf 2})}}^{(1,n)}$. Thus, the theoretical ionisation energy $J^{({\rm 1,n})}_{{\rm theo({\bf 2})}}$ is given by
\begin{equation}
\label{122}
J^{(1,{\rm n})}_{{\rm theo({\bf 2})}} = Mc^2 + E_{{\rm theo({\bf 1})}}^{(1)} - E_{{\rm theo({\bf 2})}}^{(1,{\rm n})} \,.
\end{equation}
\indent Putting now $n=1$ and $n=2$ , and subtracting the corresponding equations results in the desired energy difference $\Delta_{{\rm theo}}E_{1-2}$:
\begin{equation}
\label{123}
\Delta_{{\rm theo}}E_{1-2} = J^{({\rm 1,1})}_{{\rm theo({\bf 2})}} - J^{({\rm 1,2})}_{{\rm theo({\bf 2})}} \;.
\end{equation}
Here the {\em two\bs particle} ionisation energies $J^{({\rm 1,1})}_{{\rm theo({\bf 2})}}$ and $J^{({\rm 1,2})}_{{\rm theo({\bf 2})}}$ may be taken from the various theoretical approaches in the literature \cite{a5,a6}, whereas for the {\em one\bs particle} energies $E_{{\rm theo({\bf 1})}}^{({\rm 1})}$ \rf{122} the generally accepted values of ref. \cite{a1} are suitable. \\
\indent It must be stressed that we restrict ourselves to a {\em point\bs like} and {\em fixed} nucleus \kl of charge number $2\leq\zex\leq 100$\kr~so that the effects due to mass polarisation and nuclear size are neglected. Therefore, only the other contributions to the Lamb shifts of the one\bs and two\bs electron configurations are taken into account. This can be made more concrete by representing the energy difference $\Delta_{{\rm theo}}E_{1-2}$ \rf{123} as the sum of its ``semiclassical'' part $\Delta_{{\rm theo}}\overcirc{E}_{1-2}$ and the self\bs energy correction $\delta_sE_{1-2}$
\begin{equation}
\label{124}
\Delta_{{\rm theo}}E_{1-2} = \Delta_{{\rm theo}}\overcirc{E}_{1-2} + \delta_sE_{1-2} \;.
\end{equation}
Here, the semiclassical contribution $\Delta_{{\rm theo}}\overcirc{E}_{1-2}$ is given in terms of the semi\bs classical ionisation energies $\overcirc{J}^{({\rm 1,n})}_{{\rm theo({\bf 2})}}$ as
\begin{equation}
\label{125}
\Delta_{{\rm theo}}\overcirc{E}_{1-2} = \overcirc{J}^{({\rm 1,1})}_{{\rm theo({\bf 2})}} - \overcirc{J}^{({\rm 1,2})}_{{\rm theo({\bf 2})}}\;,
\end{equation}
\noindent where the ionization energies themselves are linked to the one- and two-particle energy levels $E^{(1)}_{{\rm theo}({\bf 1})}$ and $E^{(1,n)}_{{\rm theo}({\bf 2})}$ through
\begin{equation}
\label{126}
\overcirc{J}^{({\rm 1,n})}_{{\rm theo({\bf 2})}} = \left[ Mc^2 + \overcirc E^{(1)}_{{\rm theo}({\bf 1})}\right] - \overcirc E^{(1,n)}_{{\rm theo}({\bf 2})} \;.
\end{equation}
The energy eigenvalues $M_{*}^{(n)}c^2$ of the one\bs particle Dirac\bs Coulomb problem can be looked up in any textbook about relativistic quantum mechanics, e.g. \cite{b11}:
\begin{equation}
\label{127}
\overcirc E^{({\rm n})}_{{\rm theo}({\bf 1})} \equiv M_{*}^{(n)}c^2 = \frac{Mc^2}{\sqrt{\;1 + \frac{(\zex\aS)^2}{\lk n-1+\sqrt{1 - \lk\zex\aS\rk^2}\rk^2}}} \;.
\end{equation}
Accordingly, the self\bs energy correction $\delta_sE_{1-2}$ in the standard approaches is composed of the QED corrections \kl Lamb shift\kr~ of the ionisation energies:
\begin{equation}
\label{128}
\delta_sE_{1-2} = \delta_s J^{(1,1)}_{{\rm theo}({\bf 2})} - \delta_s J^{(1,2)}_{{\rm theo}({\bf 2})}  \;.
\end{equation}
The conventional Lamb shifts $\delta_s J^{(1,2)}_{{\rm theo}({\bf 2})}$ of the ionisation energies, with and without mass polarisation and nuclear size effects, are tabulated in ref. \cite{a6}, so that a detailed comparison with the corresponding RST predictions becomes now feasible.

\subsection{Semiclassical Results}
For the semiclassical approach, one neglects in the conventional theories the Lamb shift
which in RST corresponds to putting the self\bs interaction parameter~$u$ to zero. Under these presumptions, Table~I
presents a comparison of the theoretical predictions for the semiclassical energy difference~$\overcirc\Delta
E_{1-2}$ due to the relativistic 1/Z-expansion method \cite{a6} and RST. Here the RST
procedure consists in solving numerically the mass eigenvalue problem of Sect.~\ref{s10}
and taking the value of the energy functional~$E_T$~~(\ref{11n39}) upon the numerical
solutions.\\
\indent The experimental values~$\Delta_{{\rm exp}} E_{1-2}$~\cite{m1} are also displayed in Table~I, though it
is clear that their coincidence with the theoretical predictions can be achieved only
after including the self-energy effects (see below). In order to facilitate the analysis
of the relationship between theory and experiment, the numerical content of table~I is
illustrated by fig.1 through displaying the relative deviations ($\overcirc{\Delta}$) from the $1/Z$ expansion method:
\begin{subequations}
\begin{align}
\stackrel{\;\;_\circ}{\Delta}_{{\rm RST}} &= \frac{\Delta_{{\rm RST}}\stackrel{\;\;_\circ}{E}_{1-2} - \Delta_{1/Z}\stackrel{\;\;_\circ}{E}_{1-2}}{\Delta_{1/Z}\stackrel{\;\;_\circ}{E}_{1-2}}\label{129a} \\
\stackrel{\;\;_\circ}{\Delta}_{{\rm exp}} &= \frac{\Delta_{{\rm exp}}E_{1-2} - \Delta_{1/Z}\stackrel{\;\;_\circ}{E}_{1-2}}{\Delta_{1/Z}\stackrel{\;\;_\circ}{E}_{1-2}} \;.\label{129b}
\end{align}
\end{subequations}
\indent The semiclassical RST predictions $\Delta_{{\rm RST}}\stackrel{\;\;_\circ}{E}_{1-2}$ ($7$th column of Table I) deviate from the experimental values $\Delta_{{\rm exp}} E_{1-2}$ by $1,4 \%$ for $z_{ex}=10$ (neon) and decrease down to $0,2 \%$ for $z_{ex}=42$ (molybdenum). This effect of improvement of the RST predictions for increasing nuclear charge number $z_{ex}$ is observed also in other situations, see ref.\cite{a43}; but in view of the more precise $1/Z$-predictions ($4$th column of Table I) this result requires some explanation. The point here is that the energy difference $\Delta_{{\rm RST}}\stackrel{\;\;_\circ}{E}_{1-2}$ (\ref{125}) is composed of the two ionization energies $\stackrel{\;\;_\circ}{J}^{(1,1)}_{{\rm RST}({\bf 2})}$ and $\stackrel{\;\;_\circ}{J}^{(1,2)}_{{\rm RST}({\bf 2})}$ which therefore both do influence the numerical value for $\Delta_{{\rm RST}}\stackrel{\;\;_\circ}{E}_{1-2}$. Here, the ground-state ionization energy $\stackrel{\;\;_\circ}{J}^{(1,1)}_{{\rm RST}({\bf 2})}$ is calculated exactly and is found to be in sufficient agreement with the corresponding $1/Z$-predictions $\stackrel{\;\;_\circ}{J}^{(1,1)}_{1/Z({\bf 2})}$, see ref.\cite{a43}. However, the ionization energy 
$\stackrel{\;\;_\circ}{J}^{(1,2)}_{{\rm RST}({\bf 2})}$ of the first excited state $1s\;2s\ {}^1S_0$ is calculated by means of the spherically symmetric approximation (see the remarks above the eigenvalue equations (\ref{101})-(\ref{104})); and this approximation introduces an additional (albeit artificial) deviation which mainly is responsible for the relatively large deviations of the semiclassical RST predictions from the experimental values.\\

\begin{center}
\bf Table I
\end{center}

\indent This effect may be exemplified numerically, e.g., for nuclear charge number $z_{ex}=30$ (zinc), cf. Table I. Here the $1/Z$ prediction amounts to $\Delta_{1/Z}\stackrel{\;\;_\circ}{E}_{1-2}=8\ 955.4$ eV and thus deviates from the experimental value $\Delta_{{\rm exp}}E_{1-2}=8\ 950.2$ eV by $0.06 \%$. In contrast to this, the corresponding RST prediction is $\Delta_{{\rm RST}}\stackrel{\;\;_\circ}{E}_{1-2}=8\ 912.2$ eV and therefore has a deviation of $0.4\%$. However, this relatively large deviation is almost completely due to the ionization energy $\stackrel{\;\;_\circ}{J}^{(1,2)}_{{\rm RST}({\bf 2})}=2\ 957.5$ eV of the excited state, which deviates from the corresponding $1/Z$ value $\stackrel{\;\;_\circ}{J}^{(1,2)}_{1/Z({\bf 2})}=2\ 915.67$ eV by $1.4\%$. On the other hand, the RST prediction for the ionization energy (\ref{126}) of the groundstate is $\stackrel{\;\;_\circ}{J}^{(1,1)}_{{\rm RST}({\bf 2})}=11\ 869.7$ eV and therefore deviates from the corresponding $1/Z$ value $\stackrel{\;\;_\circ}{J}^{(1,1)}_{1/Z({\bf 2})}=11\ 871.03$ eV by only $0.01\%$. Consequently, it is just the relatively inaccurate RST prediction $\stackrel{\;\;_\circ}{J}^{(1,2)}_{{\rm RST}({\bf 2})}$ for the excited state (being caused by the spherically symmetric approximation) which is responsible for the large RST deviation from the experimental value. This result concretizes the necessity of developping some methods to solve exactly the \textit{two-dimensional} RST eigenvalue problems in order to avoid the use of the spherically symmetric approximation. Alternatively, when one wishes to stick to the spherically symmetric approximation, one might also look for a better class of trial functions than those given by equations (\ref{73})-(\ref{76}).

\begin{center}
\bf fig.1
\end{center}

\subsection{Self-Interactions}

The inclusion of the electronic self-energy works somewhat different in RST as in the
conventional theory: The RST self-interactions constitute an integral part of the theory,
namely in specifying the fibre metric~$K_{\alpha\beta}$~(\ref{913}) which must be fixed in
any way, no matter whether one wants to include or neglect the self-interactions. In contrast to this, the self-interactions of the conventional theory are added by hand to the
semiclassical results in form of perturbative corrections due to quantum
electrodynamics. The latter theory is a rather different framework which however is not able
to yield immediately the desired energy levels of atoms and molecules, so that one is
forced to resort to a two-track procedure (i.e. calculating semiclassical results and adding QED corrections). A
typical example for this is the relativistic 1/Z-expansion method of ref. \cite{a6} whose
QED corrections have been used also for the all-order technique in relativistic many-body
perturbation theory \cite{a5}.\\
\indent Thus, it becomes interesting to oppose these rather different
approaches to each other by comparing the QED-corrected semiclassical predictions
$\Delta_{1/Z}\;E_{1-2}$ of ref.s \cite{a6,a5} to the analogous RST results
$\Delta_{{\rm RST}}\;E_{1-2}$ with non-zero self-interaction parameter~$u$. Here the precise
value of~$u$ is adopted as~$u=0.03052$, because for this value of~$u$ the electronic
ground-state interaction energies and ionization energies have been found to coincide
excellently with the experimental facts for larger $z_{ex}$, see ref.s \cite{a42,a43}. Clearly one wishes to know
whether the self-interaction parameter~$u$ must adopt a universal value, or whether perhaps
it depends upon the quantum state to be considered. Table~II and fig.2 present a
comparison of the RST and 1/Z~predictions~$\Delta_{{\rm RST}}E_{1-2}$ and ~$\Delta_{1/Z}E_{1-2}$
with inclusion of the self-energy effects. The~$1/Z$ data are directly taken over from
ref. \cite{a6} with neglection of mass polarization and nuclear size, and the corresponding
RST data are obtained by numerically solving the full RST mass-eigenvalue
system~\rf{101}-\rf{1015} with~$u=0.03052$.\\
\indent Concerning the QED corrected $1/Z$ predictions $\Delta_{1/Z}E_{1-2}$ (\cite{a6}) (fourth column of Table II), the inclusion of the QED self-interactions yields an acceptable agreement with the experimental data up to medium nuclear charge numbers ($z_{ex}\lesssim 50$), see also fig.2. Here the inclusion of the QED self-energy actually generates a considerable improvement. In contrast to this, the inclusion of the RST self-interactions (by taking $u=0.03052$) results in an ambiguous shift of the energy difference $\Delta_{{\rm RST}}E_{1-2}$ (\rf{123}), see seventh column of Table II. For lower $z_{ex}$ ($<28$), the inclusion of the RST self-interactions shifts the RST predictions $\Delta_{{\rm RST}}E_{1-2}$ (\rf{123}) closer to the experimental data, whereas for the higher $z_{ex}$ ($>28$) the corresponding RST predictions move apart from the experimental values. However, the semiclassical tendency of better appoximation for the higher nuclear charge numbers does persist also after inclusion of the self-interactions.
\begin{center}
\bf fig.2
\end{center}
\indent In this sense, RST seems to represent some kind of high-energy approximation to the real world. Indeed, this feature of RST is even more manifest for the ground-state ionization energies of the helium-like ions, for which the RST predictions are more precise than the predictions of the other theoretical approaches in the literature \cite{a43}. Clearly, such a result hints at the unexhausted potentiality of RST, where it seems possible to further improve the RST predictions by finding approximations to the exact solutions which supersede the present trial functions in combination with the variational technique. 
\begin{center}
\bf Table II
\end{center}

\pagebreak
\bibliography{bibdb_2}

\begin{figure*}
\renewcommand{\arraystretch}{0.1}
{\footnotesize
\begin{tabular}{|r||r@{.}l|r@{.}l|r@{.}l||r@{.}l|r@{.}l|r@{.}l||r@{.}l|}\hline
& \multicolumn{2}{c|}{$\stackrel{\;\;_\circ}{J}^{(1,1)}_{1/Z({\bf 2})}$} & \multicolumn{2}{c|}{$\stackrel{\;\;_\circ}{J}^{(1,2)}_{1/Z({\bf 2})}$} & \multicolumn{2}{c||}{$\Delta_{1/Z}\!\stackrel{\;\;_\circ}{E}_{1-2}$} & \multicolumn{2}{c|}{$\stackrel{\;\;_\circ}{E}^{(1,2)}_{{\rm RST}({\bf 2})}$} & \multicolumn{2}{c|}{$\stackrel{\;\;_\circ}{E}^{(1,1)}_{{\rm RST}({\bf 2})}$} & \multicolumn{2}{c||}{$\Delta_{RST}\! \stackrel{\;\;_\circ}{E}_{1-2}$} & \multicolumn{2}{c|}{$\Delta_{exp}\! E_{1-2}$}\\
\raisebox{1.0ex}[-1.0ex]{$z_{\rm ex}$} & \multicolumn{2}{c|}{\cite{a6}} & \multicolumn{2}{c|}{\cite{a6}} & \multicolumn{2}{c||}{(\ref{125})} & \multicolumn{2}{c|}{$u=0 $} & \multicolumn{2}{c|}{$u=0$} & \multicolumn{2}{c||}{(\ref{125})} & \multicolumn{2}{c|}{\cite{m1}} \\\hline\hline
2   &     24 & 6 &     4 & 0 &     20 & 6 &        -&-  & 1021920 & 0 &      -&-  &   20& 6 \\
3   &     75 & 6 &    14 & 7 &     60 & 9 &        -&-  & 1021800 & 9 &      -&-  &   60& 9 \\
4   &    153 & 9 &    32 & 2 &    121 & 7 &        -&-  & 1021627 & 4 &      -&-  &  121& 7 \\
5   &    259 & 4 &    56 & 6 &    202 & 8 &        -&-  & 1021399 & 4 &      -&-  &  202& 8 \\
6   &    392 & 1 &    87 & 7 &    304 & 4 &        -&-  & 1021116 & 9 &      -&-  &  304& 4 \\
7   &    552 & 1 &   125 & 7 &    426 & 4 &        -&-  & 1020779 & 8 &      -&-  &  426& 4 \\
8   &    739 & 4 &   170 & 4 &    569 & 0 &        -&-  & 1023881 & 3 &      -&-  &  568& 9 \\
9   &    954 & 0 &   222 & 1 &    731 & 9 &        -&-  & 1019941 & 8 &      -&-  &  731& 9 \\
10  &   1195 & 9 &   280 & 5 &    915 & 4 & 1020343 & 0 & 1019440 & 7 &   902 & 3 &  915& 3 \\
12  &   1762 & 1 &   418 & 0 &   1344 & 1 & 1019602 & 1 & 1018274 & 0 &  1328 & 1 & 1343& 8 \\
13  &   2086 & 3 &   497 & 1 &   1589 & 2 & 1019180 & 0 & 1017608 & 2 &  1571 & 8 & 1589& 0 \\
14  &   2438 & 1 &   583 & 0 &   1855 & 1 & 1018723 & 4 & 1016887 & 2 &  1836 & 2 & 1854& 7 \\
15  &   2817 & 5 &   675 & 9 &   2141 & 5 & 1018232 & 3 & 1016111 & 1 &  2121 & 2 & 2141& 1 \\
17  &   3659 & 2 &   882 & 5 &   2776 & 7 & 1017146 & 0 & 1014392 & 6 &  2753 & 4 & 2776& 1 \\
18  &   4121 & 7 &   996 & 3 &   3125 & 4 & 1016550 & 7 & 1013449 & 9 &  3100 & 8 & 3124& 5 \\
20  &   5130 & 4 &  1244 & 8 &   3885 & 6 & 1015255 & 0 & 1011397 & 1 &  3857 & 9 & 3884& 2 \\
22  &   6251 & 1 &  1521 & 6 &   4729 & 5 & 1013818 & 7 & 1009119 & 8 &  4698 & 9 & 4727& 7 \\
23  &   6853 & 8 &  1670 & 6 &   5183 & 2 & 1013047 & 5 & 1007896 & 5 &  5151 & 0 & 5181& 0 \\
25  &   8144 & 1 &  1990 & 1 &   6154 & 0 & 1011398 & 2 & 1005279 & 5 &  6118 & 7 & 6151& 1 \\
27  &   9548 & 4 &  2338 & 4 &   7210 & 0 & 1009605 & 5 & 1002433 & 9 &  7171 & 6 & 7206& 3 \\
28  &  10293 & 7 &  2523 & 5 &   7770 & 2 & 1008655 & 0 & 1000924 & 8 &  7730 & 2 & 7766& 0 \\
30  &  11871 & 0 &  2915 & 7 &   8955 & 4 & 1006644 & 8 &  997732 & 6 &  8912 & 2 & 8950& 2 \\
32  &  13565 & 0 &  3337 & 5 &  10227 & 5 & 1004487 & 6 &  994306 & 5 & 10181 & 1 &10220& 0 \\
33  &  14456 & 1 &  3559 & 6 &  10896 & 5 & 1003353 & 5 &  992505 & 1 & 10848 & 4 &10890& 5 \\
35  &  16327 & 3 &  4026 & 5 &  12300 & 8 & 1000973 & 0 &  988723 & 8 & 12249 & 2 &12293& 1 \\
36  &  17307 & 7 &  4271 & 4 &  13036 & 3 &  999726 & 1 &  986743 & 2 & 12982 & 9 &13027& 0 \\
37  &  18318 & 3 &  4524 & 1 &  13794 & 2 &  998441 & 3 &  984702 & 2 & 13739 & 1 &    -&-  \\
38  &  19359 & 2 &  4784 & 5 &  14574 & 7 &  997118 & 2 &  982600 & 4 & 14517 & 8 &    -&-  \\
40  &  21532 & 6 &  5328 & 8 &  16203 & 8 &  994356 & 0 &  978212 & 8 & 16143 & 2 &    -&-  \\
42  &  23829 & 7 &  5905 & 0 &  17924 & 8 &  991437 & 6 &  973577 & 2 & 17860 & 4 &17908& 4 \\
45  &  27511 & 1 &  6830 & 1 &  20681 & 0 &  986761 & 5 &  966150 & 7 & 20610 & 8 &    -&-  \\
50  &  34291 & 8 &  8539 & 5 &  25752 & 3 &  978150 & 3 &  952478 & 7 & 25671 & 6 &    -&-  \\
54  &  40317 & 1 & 10064 & 2 &  30252 & 9 &  970498 & 0 &  940335 & 0 & 30163 & 0 &    -&-  \\
60  &  50405 & 6 & 12628 & 6 &  37777 & 0 &  957681 & 2 &  920009 & 2 & 37672 & 0 &    -&-  \\
65  &  59830 & 0 & 15037 & 2 &  44792 & 8 &  945700 & 4 &  901026 & 9 & 44673 & 5 &    -&-  \\
70  &  70242 & 4 & 17712 & 7 &  52529 & 7 &  932452 & 8 &  880058 & 5 & 52394 & 3 &    -&-  \\
75  &  81714 & 9 & 20678 & 2 &  61036 & 7 &  917840 & 7 &  856957 & 9 & 60882 & 8 &    -&-  \\
80  &  94335 & 7 & 23961 & 9 &  70373 & 9 &  901745 & 5 &  831546 & 6 & 70198 & 9 &    -&-  \\
\hline\hline
                                        &     \multicolumn{14}{c|}{{\it all data in}\/ eV}                                              \\ \hline

\end{tabular}
}\\
\flushleft{{\large {\bf Table I}:\quad{\it Semiclassical Energy Differences $\Delta \stackrel{\;\;_\circ}{E}_{1-2}$ (\ref{121}),(\ref{125}) }}}\\
\qquad The semiclassical results of the $1/Z$ expansion method \cite{a6} (fourth column) for the excitation energy $\Delta \stackrel{\;\;_\circ}{E}_{1-2}$ are compared to the present RST predictions $\Delta_{RST}\! \stackrel{\;\;_\circ}{E}_{1-2}$ according to equation (\ref{125}). The relatively large deviations of the RST results(seventh column) from the experimental values (last column) are attributed to the anisotropy effect, which is neglected by the spherically symmetric approximation. The experimental values are taken from reference \cite{m1}.
\end{figure*}

\begin{figure*}
\renewcommand{\arraystretch}{0.1}
{\footnotesize
\begin{tabular}{|r||r@{.}l|r@{.}l|r@{.}l||r@{.}l|r@{.}l|r@{.}l||r@{.}l|}\hline
& \multicolumn{2}{c|}{$J^{(1,1)}_{1/Z({\bf 2})}$} & \multicolumn{2}{c|}{$J^{(1,2)}_{1/Z({\bf 2})}$} & \multicolumn{2}{c||}{$\Delta_{1/Z}\!E_{1-2}$} & \multicolumn{2}{c|}{$J^{(1,1)}_{{\rm RST}({\bf 2})}$} & \multicolumn{2}{c|}{$J^{(1,2)}_{{\rm RST}({\bf 2})}$} & \multicolumn{2}{c||}{$\Delta_{RST}\!E_{1-2}$} & \multicolumn{2}{c|}{$\Delta_{exp}\! E_{1-2}$}\\
\raisebox{1.0ex}[-1.0ex]{$z_{\rm ex}$} & \multicolumn{2}{c|}{\cite{a6}} & \multicolumn{2}{c|}{\cite{a6}} & \multicolumn{2}{c||}{(\ref{123})} & \multicolumn{2}{c|}{$u=0.03052 $} & \multicolumn{2}{c|}{$u=0.03052$} & \multicolumn{2}{c||}{(\ref{125})} & \multicolumn{2}{c|}{\cite{m1}} \\\hline\hline
 2 &    24 & 6 &     4 & 0 &    20 & 6 &      -&-  &      -&-  &      -&-  &    20 & 6 \\
 3 &    75 & 6 &    14 & 7 &    60 & 9 &    74 & 5 &      -&-  &      -&-  &    60 & 9 \\
 4 &   153 & 9 &    32 & 2 &   121 & 7 &   152 & 7 &      -&-  &      -&-  &   121 & 7 \\
 5 &   259 & 4 &    56 & 6 &   202 & 8 &   258 & 2 &      -&-  &      -&-  &   202 & 8 \\
 6 &   392 & 1 &    87 & 7 &   304 & 4 &   390 & 9 &      -&-  &      -&-  &   304 & 4 \\
 7 &   552 & 1 &   125 & 7 &   426 & 4 &   550 & 9 &      -&-  &      -&-  &   426 & 4 \\
 8 &   739 & 3 &   170 & 4 &   568 & 9 &   738 & 2 &      -&-  &      -&-  &   568 & 9 \\
 9 &   953 & 9 &   222 & 0 &   731 & 9 &   952 & 8 &      -&-  &      -&-  &   731 & 9 \\
10 &  1195 & 8 &   280 & 5 &   915 & 3 &  1194 & 8 &   290 & 8 &   904 & 0 &   915 & 3 \\ 
12 &  1761 & 8 &   418 & 0 &  1343 & 8 &  1760 & 9 &   431 & 0 &  1330 & 0 &  1343 & 8 \\
13 &  2086 & 0 &   497 & 0 &  1589 & 0 &  2085 & 2 &   511 & 4 &  1573 & 8 &  1589 & 0 \\
14 &  2437 & 7 &   583 & 0 &  1854 & 7 &  2436 & 9 &   598 & 7 &  1838 & 2 &  1854 & 7 \\
15 &  2816 & 9 &   675 & 8 &  2141 & 1 &  2816 & 3 &   693 & 0 &  2123 & 3 &  2141 & 1 \\
17 &  3658 & 3 &   882 & 4 &  2775 & 9 &  3658 & 0 &   902 & 5 &  2755 & 6 &  2776 & 1 \\
18 &  4120 & 7 &   996 & 1 &  3124 & 6 &  4120 & 6 &  1017 & 7 &  3102 & 9 &  3124 & 5 \\
20 &  5128 & 9 &  1244 & 6 &  3884 & 3 &  5129 & 3 &  1269 & 4 &  3859 & 9 &  3884 & 2 \\
22 &  6249 & 0 &  1521 & 3 &  4727 & 7 &  6249 & 7 &  1549 & 4 &  4700 & 3 &  4727 & 7 \\
23 &  6851 & 3 &  1670 & 3 &  5181 & 0 &  6852 & 7 &  1700 & 1 &  5152 & 6 &  5181 & 0 \\
25 &  8140 & 8 &  1989 & 7 &  6151 & 1 &  8143 & 1 &  2023 & 2 &  6119 & 9 &  6151 & 1 \\
27 &  9544 & 2 &  2337 & 9 &  7206 & 3 &  9547 & 5 &  2375 & 2 &  7172 & 3 &  7206 & 3 \\
28 & 10288 & 9 &  2522 & 9 &  7766 & 0 & 10293 & 0 &  2562 & 3 &  7730 & 7 &  7766 & 0 \\
30 & 11865 & 0 &  2914 & 8 &  8950 & 2 & 11870 & 4 &  2958 & 5 &  8911 & 9 &  8950 & 2 \\
32 & 13557 & 5 &  3336 & 4 & 10221 & 1 & 13564 & 4 &  3384 & 6 & 10179 & 8 & 10220 & 0 \\
33 & 14447 & 8 &  3558 & 4 & 10889 & 4 & 14455 & 6 &  3609 & 0 & 10846 & 6 & 10890 & 5 \\
35 & 16317 & 1 &  4025 & 1 & 12292 & 0 & 16356 & 9 &  4080 & 7 & 12246 & 2 & 12293 & 1 \\
36 & 17296 & 6 &  4269 & 9 & 13026 & 7 & 17307 & 5 &  4328 & 1 & 12979 & 4 & 13027 & 0 \\ 
37 & 18306 & 1 &  4522 & 3 & 13784 & 4 & 18318 & 1 &      -&-  &      -&-  &      -&-  \\
38 & 19345 & 8 &  4782 & 6 & 14563 & 2 & 19359 & 1 &  4846 & 3 & 14512 & 8 &      -&-  \\
40 & 21516 & 8 &  5326 & 6 & 16190 & 2 & 21532 & 7 &  5396 & 1 & 16136 & 6 &      -&-  \\
42 & 23811 & 1 &  5902 & 3 & 17908 & 8 & 23830 & 0 &  5978 & 0 & 17852 & 0 & 17908 & 4 \\
45 & 27487 & 6 &  6826 & 7 & 20660 & 9 & 27511 & 8 &  6912 & 6 & 20599 & 2 &      -&-  \\
50 & 34258 & 5 &  8534 & 6 & 25723 & 9 & 34293 & 3 &  8639 & 7 & 25653 & 6 &      -&-  \\
54 & 40274 & 1 & 10057 & 6 & 30216 & 5 & 40319 & 4 & 10180 & 4 & 30139 & 0 &      -&-  \\
60 & 50344 & 5 & 12619 & 1 & 37725 & 4 & 50409 & 4 & 12772 & 5 & 37636 & 9 &      -&-  \\
65 & 59750 & 1 & 15024 & 4 & 44725 & 7 & 59835 & 1 & 15207 & 9 & 44627 & 2 &      -&-  \\ 
70 & 70139 & 9 & 17695 & 9 & 52444 & 0 & 70249 & 3 & 17914 & 0 & 52335 & 3 &      -&-  \\
75 & 81585 & 5 & 20656 & 4 & 60929 & 1 & 81724 & 0 & 20914 & 4 & 60809 & 6 &      -&-  \\
80 & 94174 & 3 & 23933 & 9 & 70240 & 4 & 94347 & 4 & 24237 & 5 & 70109 & 9 &      -&-  \\ \hline\hline
                                        &     \multicolumn{14}{c|}{{\it all data in}\/ eV}                                              \\ \hline

\end{tabular}
}\\
\flushleft{{\large {\bf Table II}:\quad{\it Energy Differences $\Delta E_{1-2}$ with Self-Interaction}}}\\
\qquad The semiclassical predictions for the energy difference $\Delta E_{1-2}$ become improved considerably for the conventional $1/Z$ expansion method but only partly for RST by including the self-energy effects. The moderate precision of the RST predictions is a consequence of the use of (i) the spherically symmetric approximation and (ii) the simple one-particle trial functions \rf{73}\bs\rf{76}. It is only the use of these very rough approximations which devaluates the RST predictions down to the level of the simple Hartree-Fock approach.
\end{figure*}

\setlength{\unitlength}{1cm}
\begin{figure*}
\epsfig{file=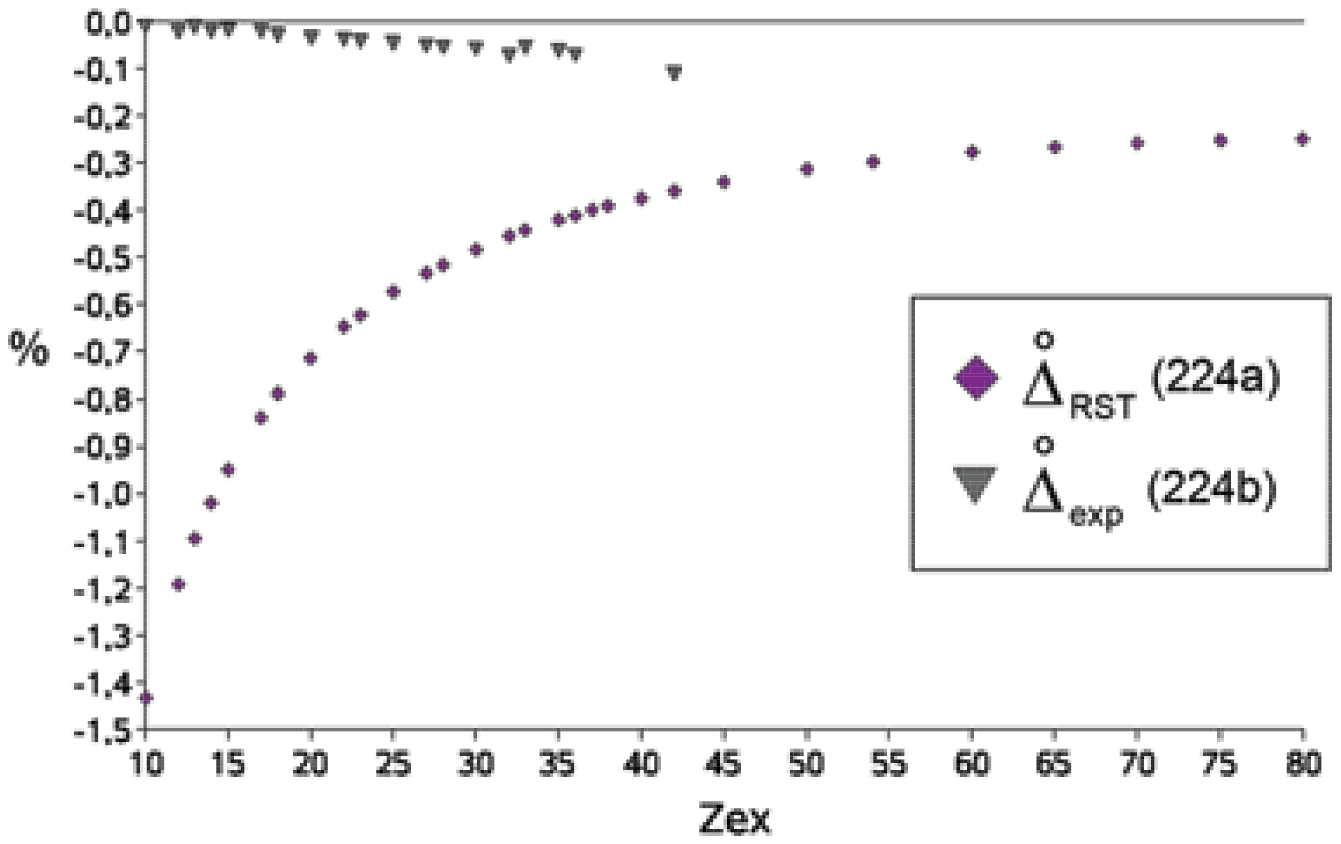,width=15cm}
\flushleft{{\large {\bf Fig.1}:\quad{\it Relative Deviations $\stackrel{\;\;_\circ}{\Delta}$ (\ref{129a})-(\ref{129b})}}}\\
\qquad The relative semiclassical deviations from the experimental values ($\blacktriangledown$) are larger for the RST approach ($\blacklozenge$) than for the $1/Z$ expansion method \cite{a6} because the spherically symmetric approximation of RST induces an extra (albeit artificial) deviation in addition to the missing of the correlation energy beyond Hartree-Fock. For high nuclear charge numbers $z_{ex}$, the RST predictions ($\blacklozenge$) approximate the $1/Z$ predictions (horizontal axis). For this asymptotic ($z_{ex}>>1$) coincidence of RST and $1/Z$ predictions see ref.\cite{a43}.  
\end{figure*}

\setlength{\unitlength}{1cm}
\begin{figure*}
\epsfig{file=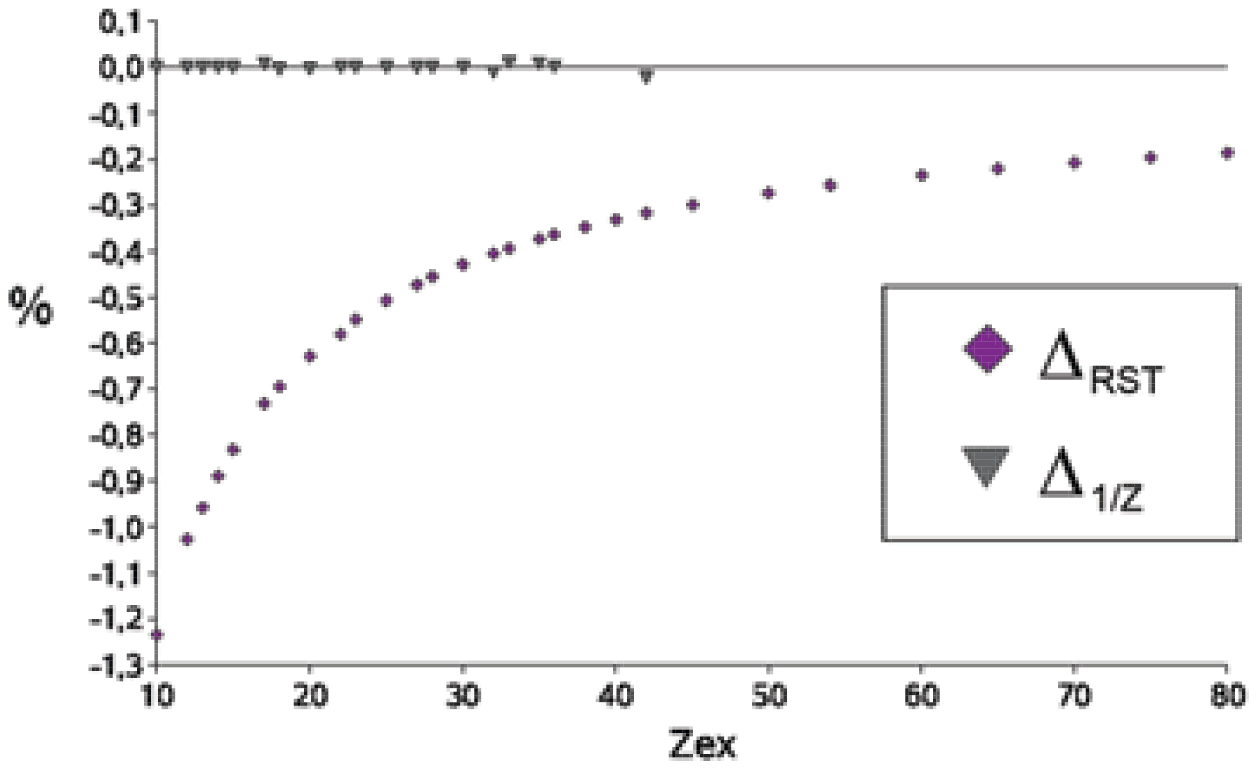,width=15cm}
\flushleft{{\large {\bf Fig.2}:\quad{\it Relative Deviations (\ref{129a})-(\ref{129b}) with Self-Interactions}}}\\
\qquad The inclusion of the QED corrections yields a considerable improvement of the $1/Z$ predictions ($\blacktriangledown$), see fig.1; whereas the modification of the RST predictions ($\blacklozenge$) is ambiguous: improvement for low $z_{ex}$($<28$) but deterioration for the higher $z_{ex}$ ($>28$). This is interpreted as an artifical effect due to the use of the spherically symmetric approximation in combination with the simple one\bs particle trial functions \rf{73}\bs\rf{76}.
\end{figure*}

\end{document}